\documentclass[12pt]{article}
\usepackage{epsfig,equations,graphics,amssymb}
\usepackage{cite}

\begin{document}

\def\nicefrac#1#2{\hbox{${#1\over #2}$}}
\textwidth 10cm
\setlength{\textwidth}{13.7cm}
\setlength{\textheight}{23cm}

\oddsidemargin 1cm
\evensidemargin -0.2cm
\addtolength{\topmargin}{-2.5 cm}

\newcommand{\nn}{\nonumber}
\newcommand{\raw}{\rightarrow}
\newcommand{\be}{\begin{equation}}
\newcommand{\ee}{\end{equation}}
\newcommand{\bea}{\begin{eqnarray}}
\newcommand{\eea}{\end{eqnarray}}
\newcommand{\dl}{\stackrel{\leftarrow}{D}}
\newcommand{\dr}{\stackrel{\rightarrow}{D}}
\newcommand{\dd}{\displaystyle}
\newcommand{\Ln}{{\rm Ln}}

\pagestyle{empty}
 
\begin{flushright}
FTUAM 00/20 \\
IFT-UAM/CSIC 00/42  \\
KA-TP-10-2001\\
hep--ph/0105097 \\
March 2001 \\
\end{flushright}
\vskip1cm

\renewcommand{\thefootnote}{\fnsymbol{footnote}}
\begin{center}
{\Large\bf
SUSY-QCD decoupling properties in \boldmath$H^+ \rightarrow t \bar b$ 
decay}\\[1cm]
{\large Mar\'{\i}a J. Herrero$^a$, 
Siannah Pe\~naranda$^b$~\footnote{On leave from Departamento de 
F\'{\i}sica Te\'{o}rica, Universidad Aut\'{o}noma de Madrid, Madrid, 
Spain.} and 
David Temes$^a$~\footnote{electronic addresses:
herrero@delta.ft.uam.es, siannah@particle.uni-karlsruhe.de, 
temes@delta.ft.uam.es}
}\\[6pt]
{\it $^a$ Departamento de F\'{\i}sica Te\'{o}rica \\
   Universidad Aut\'{o}noma de Madrid,
   Cantoblanco, 28049 Madrid, Spain. \\
$^b$ Institut f\"ur Theoretische Physik \\
   Universit\"at Karlsruhe, Kaiserstra\ss e 12, D-76128 Karlsruhe.}
\\[1cm]

\begin{abstract}
\vspace*{0.4cm}
The SUSY-QCD radiative corrections to the $\Gamma (H^+ \rightarrow t \bar b)$
partial decay width are analyzed
within the Minimal Supersymmetric Standard Model at the one-loop
level, ${\mathcal O}(\alpha_s)$, and in the decoupling limit. 
We present the analytical expressions of these corrections 
in the large SUSY masses limit and study the
decoupling behaviour of these co\-rrections in various limiting
cases. We find that if the SUSY mass parameters are large and of the
same order, the one loop SUSY-QCD 
corrections {\it do not decouple}. The non-decoupling contribution is
enhanced by $\tan \beta$ and therefore large corrections are expected in the large
$\tan \beta$ limit. In contrast, we also find that the SUSY-QCD corrections
decouple if the masses of either the squarks or the gluinos are
separately taken large. 
\end{abstract}

\end{center}

\vfill
\clearpage

\renewcommand{\thefootnote}{\arabic{footnote}}
\setcounter{footnote}{0}

\pagestyle{plain}

\section{Introduction} 

Despite its tremendous success in the agreement between the Standard
Model (SM) predictions and the experimental data, this model 
contains a variety of theoretical pro\-blems which cannot be solved 
without the introduction
of some new physics. Here we will be concerned about two candidates for physics
beyond the SM: Supersymmetry (SUSY) and Extended Higgs Sectors.

The minimal extension of the SM Higgs
sector is a two Higgs doublet model
(2HDM). This model predicts three neutral
Higgs bosons, two CP even ($h^{o}$, $H^{o}$) one 
CP odd ($A^{o}$), and two charged bosons
($H^{\pm}$); and includes one parameter, $\tan
\beta$, defined as the ratio of the two Higgs vacuum
expectation values ({\em vevs})~\cite{MSSMHiggssector}. On the other hand, 
cons\-truc\-ting a Minimal
Supersymmetric extension of the Standard Model (MSSM)~\cite{HaberKane} implies
a two Higgs doublet sector, with the same spectrum as in a 2HDM
extension of the SM
but with constrained Yukawa and Higgs self-couplings. To be more
specific, the MSSM Higgs sector belongs to the so-called type II
models, in which one Higgs doublet couples to u-like quarks and the
other one to d-like quarks ~\cite{MSSMHiggssector}.

Looking for non-SM
effects on the Higgs phenomeno\-lo\-gy of these extended 
models we realize that, 
while both the neutral Higgs
bosons may not be easily
distinguishable from that of the SM, the discovery of $H^{\pm}$ and the
determination of its mass and couplings are expected to be a very
clear signal of physics beyond the SM in the Higgs sector. This is the
main reason
why the search for charged Higgs bosons is one of the major tasks at
present and future colliders.

Concerning the present experimental search of charged Higgs bosons the
situation is as follows.
LEP2 has set a model independent lower limit on the $H^{+}$
mass, $m_{H^+} \geq 77.4$ GeV~\cite{ChargedHLEP}.
At Tevatron, the CDF and D0 collaborations have searched for $H^{\pm}$
bosons in top decays through the process $p\bar p \rightarrow t\bar
t$, with at least one of the top quarks decaying via $t \rightarrow 
H^{+} b$, leading to an excess of $\tau$ due to the $H^{+}
\rightarrow \tau^{+} \nu$ decay; they have excluded regions with light
$H^+$ and large values of $\tan \beta$, for example if
$M_{H^+}= 100$ GeV, $\tan \beta \geq 60$ is excluded~\cite{tevatron}. 
In any case, the allowed region in the $(\tan \beta - M_{H^\pm})$ plane can be significantly
modified by quantum corrections~\cite{RadCorrec}. 

Here we consider the case of a heavy charged Higgs boson, $M_{H^+} \geq
m_t + m_b$. In this case $H^{+}$ bosons will decay mostly into 
$t \bar b$ pairs with approximately $BR
\geq 85\%$~\cite{spiraQCD}, depending on the values of $\tan \beta$
and $M_{H^+}$. 
On the other hand the
two main $H^{+}$ production subprocesses which will provide sizeable
cross sections at LHC 
are: $g\bar b \rightarrow \bar t H^{+}$~\cite{Bawa:1990pc} 
and $gg \rightarrow \bar tH^{+}b$~\cite{Gunion:1994sv}. 
Thus, the importance of the
$H^{+} \rightarrow t\bar b$ decay comes in two directions:
it is the dominant decay over most of the parameter space and its
associated effective vertex $H^{+}\bar t b$ appears in the two main
production processes.

Once a charged Higgs Boson is found, a precise determination of its couplings
to SM particles, which are sensitive to radiative corrections, can give
us indirect information about extra new physics beyond the SM. In
parti\-cu\-lar one may wonder if the $H^{\pm}$ is itself an indirect
signal of supersymmetry and explore if its corrected couplings to SM
fermions are like in the MSSM or like in a non-SUSY 2HDM.
In order to compare the predictions in the MSSM and
in the 2HDM extension of the SM for these couplings, we assume here
the most pessimistic scenario where the
ge\-nui\-ne supersymmetric spectrum is very heavy as compared with the
electroweak scale. This situation corresponds to the decoupling of 
supersymme\-tric
particles from the rest of the MSSM spectrum, namely, the SM particles
and the MSSM Higgs sector. In this decoupling limit the SUSY
particles can not be produced directly in the
colliders and we are constrai\-ned to indirect searches for SUSY. Here
we will look for indirect heavy SUSY signals through their effect on radiative
corrections to the $H^+ \rightarrow t\bar b$ decay.

Recent works on the decoupling limit have shown that all the genuine
heavy SUSY
particles and heavy Higgs bosons ($H^{o}$, $A^{o}$ and  $H^{\pm}$) of
the MSSM
decouple, at one-loop order, from the low-energy electroweak gauge
bosons physics~\cite{TesisS}. On the other hand, concerning the Higgs
physics, the decoupling behaviour of the
one-loop SUSY-QCD radiative corrections to the $\Gamma
(h^{o} \rightarrow b\bar b)$ partial width have
been studied in~\cite{HaberTemes}. There it was found that, for fixed
masses of the extra Higgses ($H^{o}$,
$A^{o}$ and  $H^{\pm}$), the SUSY-QCD corrections from nearly heavy
degenerate gluinos and squarks do not decouple in the $\Gamma(h^{o}
\rightarrow b\bar b)$ decay width. The SUSY-QCD corrections to the
main Higgs bosons and
top decays have also been analyzed in ~\cite{RADCOR2000}.

In this paper we study the MSSM radiative corrections to the $H^{+}
\rightarrow t\bar b $ decay width at one loop level and to leading order in
$\alpha_s$, and we analyze their behaviour in the decoupling
limit of heavy SUSY particles. These corrections come from the SUSY-QCD ~\cite{SQCD,QCDandS}
and pure QCD sectors~\cite{qcdhtb} and
are known to provide the dominant contributions to the $\Gamma (H^{+}
\rightarrow t\bar b)$ partial width. The QCD corrections range from $+10
\%$ to $-50 \%$ and the
SUSY-QCD corrections can be even larger in a large region of the SUSY
parameter space. Since a 2HDM
extension of the SM has no SUSY-QCD contributions to $H^{+}
\rightarrow t\bar b $, one could use
these corrections in order to distinguish between the MSSM and the 2HDM.

Here we analyze the full diagrammatic formulae for the
on-shell one-loop SUSY-QCD corrections to the $H^+ \rightarrow t \bar b$ decay.  We
perform expansions in inverse powers of the SUSY masses in order to
examine the decoupling behavior when these masses are large
compared to the electroweak scale.  
The SUSY-QCD corrections depend on a number of
different MSSM mass parameters, and we will see that the relative
sizes of these parameters do affect qualitatively the decoupling behaviour.  To remain as
model-independent as possible, we make no assumptions about relations
among the MSSM parameters that may arise from grand unification or
specific SUSY-breaking scenarios.  We consider the soft-SUSY-breaking
parameters and the $\mu$ parameter as unknown parameters whose
magnitudes are all of order 1~TeV. We examine in great detail the case
of large $\tan\beta$, for which
the SUSY-QCD corrections are enhanced.
This enhancement gives rise to a significant one-loop correction, even for
moderate and large values of the SUSY masses.

This paper is organized as follows.
In Section~\ref{sec:masses} we define our notation and briefly review
the Higgs and squark sectors of the MSSM.
In Section~\ref{sec:exactcorr} we review the exact one loop result for
the SUSY-QCD corrections to the $H^+ \rightarrow t \bar b$ partial decay width.
In Section~\ref{sec:LargeMP} we derive analytic expressions for
these co\-rrec\-tions in the limit of large SUSY masses and for
several extreme squark mixing cases.  We also analyze
the decoupling of the SUSY-QCD corrections for various hierarchies of
mass parameters, and compare the ana\-lytic approximations to
the exact one-loop result. Finally, we summarize our conclusions in
Section~\ref{sec:conclusions}.

\section{MSSM Higgs and squark sectors}
\label{sec:masses}

In the MSSM there are two isospin Higgs doublets containing
eight degrees of freedom. After the electroweak symmetry-breaking
mecha\-nism, three of these eight degrees of freedom are absorbed by the $Z$
and $W^{\pm}$ gauge bosons, leading to the existence of five physical Higgs
particles. These consist of two CP-even neutral scalar particles
$h^{o}$, $H^{o}$, one CP-odd neutral pseudoscalar particle $A^{o}$ and two
charged scalar particles $H^{\pm}$. Due to supersy\-mme\-try, the parameters of
the Higgs sector are constrained and, at
tree-level, it turns out that the Higgs masses and mixing angle
depend on just two unknown parameters.
These are commonly chosen to be the mass
of the CP-odd neutral Higgs boson, $M_{A}$,
and the ratio of the {\em{vevs}}
of the two Higgs doublets, $\tan\beta = v_2/v_1$.
The Higgs bosons masses at tree level, in terms of these parameters, are
given by:
\bea
M^2_{H^\pm} &=&  M_A^2 + M_W^2 \nonumber\\
M^2_{H^{o}, h^{o}} &=& \frac{1}{2}
    \left[ M_A^2 + M_Z^2 \pm \sqrt{\left(M_A^2 + M_Z^2\right)^2 -
           4 M_A^2 M_Z^2 \cos^2 2 \beta }\, \right]\,.
\eea

We choose a convention where the {\em{vevs}} are positive so that $0<\beta<\pi/2$.
The mixing angle in the neutral sector is given at tree-level by:
\begin{equation}
 \tan 2 \alpha = \tan 2 \beta\, {{M_A^2 + M^2_Z}\over{M_A^2 -M^2_Z}}\,.
\end{equation}
In the conventions employed here, $-\pi/2<\alpha<0$. 

We now discuss the parameters of the third generation squark
sector. For simplicity, we assume here that there is no
intergenerational flavour mixing.
The tree-level stop and sbottom squared-mass matrices are:

\begin{equation}
{\cal M}_{\tilde{t}}^2 =\left(\begin{array}{cc}
M_L^2 &
m_t\, X_t\\ m_t\, X_t & M_R^2
\end{array} \right)\,,
\label{eq:stopmatrix}
\end{equation}

\begin{equation}
{\cal M}_{\tilde{b}}^2 =\left(\begin{array}{cc}
M_L^{'2} &
m_b\, X_b\\ m_b\, X_b &M_R^{'2} 
\end{array} \right)\,,
\label{eq:sbottommatrix}
\end{equation}
where: 
\bea 
M_L^2&=&M_{\tilde Q}^2+m_t^2+\cos{2\beta}(1/2- 2/3 s_W^2)\,M_Z^2 \nonumber \\
M_R^2&=&M_{\tilde U}^2+m_t^2+2/3 \cos{2\beta}\,s_W^2\,M_Z^2 \nonumber \\
X_t&=&A_t-\mu\cot\beta\, \nonumber \\
M_L^{'2}&=&M_{\tilde Q}^2+m_b^2-\cos{2\beta}(1/2- 1/3 s_W^2)\,M_Z^2 \nonumber \\
M_R^{'2}&=&M_{\tilde D}^2+m_b^2\,-1/3 \cos{2\beta}\,s_W^2\,M_Z^2 \nonumber \\
X_b&=&A_b-\mu\tan\beta\,,
\label{eq:MLRtb}
\eea
and $s_W\equiv \sin\theta_W$. The parameters
$M_{\tilde Q}$, $M_{\tilde D}$ and $M_{\tilde U}$ are the soft-SUSY-breaking 
masses
for the third-generation SU(2) squark doublet $\tilde Q=(\widetilde
t_L, \widetilde b_L)$
and the singlets $\tilde D=\widetilde b_R$ and $\tilde U=\widetilde t_R$, respectively.
$A_{b,t}$ are the corresponding soft-SUSY-breaking trilinear
couplings and $\mu$ is the bilinear coupling of the two Higgs doublet. 
The squarks (sbottom and stop) mass eigenstates are given by:
\begin{equation}
\left( \begin{array}{c}
\tilde q_1\\ \tilde q_2 
\end{array}\right) = (R^{(q)})^{-1} \left( \begin{array}{c}
\tilde q_L\\ \tilde q_R 
\end{array}\right)\,,
\label{eq:squarkrotation}
\end{equation}
where
\begin{equation}
R^{(q)} = \left( \begin{array}{cc}
\cos \theta_{\tilde q} & - \sin \theta_{\tilde q} \\  \sin
\theta_{\tilde q} & \cos \theta_{\tilde q}
\end{array}\right)\,. 
\label{eq:matrixrotation}
\end{equation}
The stop and sbottom mass eigenvalues are given by\footnote{Note that
in our convention, $M_{\tilde t_1}> M_{\tilde t_2}$ and 
$M_{\tilde b_1}> M_{\tilde b_2}$}:
\bea
    M^2_{\tilde t_{1,2}} = \frac{1}{2}\left[ M_L^2 + M_R^2
    \pm \sqrt{ (M_L^2 - M_R^2)^2 + 4 m_t^2 X_t^2 } \right]\,,\nonumber\\
 M^2_{\tilde b_{1,2}} = \frac{1}{2}\left[ M_L^{'2} + M_R^{'2}
    \pm \sqrt{ (M_L^{'2} - M_R^{'2})^2 + 4 m_b^2 X_b^2 } \right]\,.
\eea
And the mixing angles $\theta_{\tilde q}$ $(q=t,b)$ are given by:
\bea \label{thetab}
    \cos 2 \theta_{\tilde t} = \frac{M_L^2 - M_R^2}
    {M^2_{\tilde t_1}-M^2_{\tilde t_2}}\,&,&\,\,\,\,\,\,
    \cos 2 \theta_{\tilde b} = \frac{M_L^{'2} - M_R^{'2}}
    {M^2_{\tilde b_1}-M^2_{\tilde b_2}}, \nonumber \\
    \sin 2 \theta_{\tilde q} &=&
    \frac{2 m_q X_q}{M^2_{\tilde q_1}-M^2_{\tilde q_2}}\,.
\eea

Concerning the experimental bounds on the SUSY and Higgs masses that
are relevant for this work, we briefly summarize next the present
situation. As mentioned before, from combined searches of charged
Higgs bosons at LEP one gets a general lower limit of $M_{H^+} \geq
77.4$ GeV ~\cite{ChargedHLEP}, valid at the
$95\%$ C.L. and for any 
value of the branching ratio $B(H^+ \rightarrow \tau^+ \nu)$.
On the other hand, from absence of signals in direct searches at the 
Tevatron~\cite{Affolder:2000wp}, the sbottoms
must be heavier than about $140$ GeV, assuming that the mass of the lightest
neutralino $\tilde \chi^0_1$ is less than half the mass of the lighter
sbottom. If $m_{\tilde \chi^0_1}=40$ GeV the maximum excluded value for the 
scalar top mass is about $110$ GeV at the $95\%$ C.L..
For heavier neutralinos, the Tevatron searches lose efficiency.
In this region the
direct searches at LEP~\cite{Barate:2000tu} place a lower bound on the sbottom
masses of about $91$ GeV under the assumption of a dominant 
$\tilde b \to b \tilde \chi^0_1$ decay. For $\tilde t - \tilde \chi^0_1$
mass splittings in the range from $6$ to $40$ GeV, {\it i.e.}, a region not 
accessible to the Tevatron searches, the lower limit on scalar top mass is
$83$ GeV, independently of the $\tilde t$ mixing angle. All these lower limit 
results are at $95\%$ C.L.. Finally, the limits on the gluino mass $M_{\tilde g}$ are more model-dependent.  If one
assumes relations between the gaugino masses such that they unify at
the GUT scale, then $M_{\tilde g}$ is constrained from direct
searches at the Tevatron to be greater than 173 GeV, independently of the
squark masses~\cite{Tevatronsquarkgluino}.

\section{$H^+ \rightarrow t \bar b$ partial decay width}
\label{sec:exactcorr}

The $H^+ \rightarrow t \bar b$ partial decay width at tree level is
determined by the interaction Lagrangian describing the  $H^+ \bar t b$ 
vertex, which is given by~\cite{Gunion:1989we}:
\be
{\cal L}_{Htb}={g\,V_{tb}\over\sqrt{2}M_W}\,H^{+}\,\bar{t}\, [m_t\cot\beta\,P_L
+ m_b\tan\beta\,P_R]\,b+{\rm h.c.}\,,
\label{eq:LtbH}
\ee
where $P_{L,R}=1/2(1\mp\gamma_5)$ are the chirality projection
operators and $V_{tb}$ is the corresponding CKM matrix element. Here
we do not consider mixing between different families and we set $V_{tb}=1$.
 
In this work we are interested in the dominant radiative corrections
to the tree-level partial decay width $H^+ \rightarrow t \bar b$, which 
are known
to come from QCD loops corrections. The leading corrections at
one-loop level and to order $\alpha_s$ in the context of the MSSM were
computed in~\cite{SQCD,QCDandS,qcdhtb}. They included the pure QCD
corrections from quarks and gluons and the genuine SUSY-QCD
corrections from gluinos and squarks. The standard QCD corrections
were first computed in~\cite{qcdhtb} and can be large,
ranging from $+10\%$ to $-50\%$ with respect to the tree level
contribution. The SUSY-QCD corrections, computed by using a
diagrammatic approach in~\cite{SQCD,QCDandS}, have been found to be
comparable or even larger than the standard QCD contributions in a
large region of the SUSY parameter space.

For completeness, and since the SUSY-QCD corrections
are our starting point in order to study their behaviour in the large
SUSY mass limit, we have reproduced the results in~\cite{SQCD,QCDandS} 
and we present
in the following a short summary of the most relevant analytical
expressions.

Following the standard renormalization procedure, there are two kinds of
contributions to the partial
decay width: one coming
from the loop diagrams and another one coming from the 
counterterms. Taking this into account, we can write the 
$H^+ \rightarrow t \bar b$
partial decay width at the one-loop level and to order $\alpha_s$ in
the following way\footnote{The factor 2 used here is a convention such
that the corresponding correction to the effective coupling is:
$$
    g_{H^+ b \bar t}^{1-loop} = g_{H^+ b \bar t}^{tree}
    (1 + \Delta^{loops} + \Delta^{CT})\,.
$$}:
\begin{equation}
    {\Gamma_1}(H^+ \to t \bar b) = \Gamma_0 (H^+ \to t \bar b)
    (1 + 2 \Delta^{loops} + 2 \Delta^{CT})\,,\\
\end{equation}
where $\Gamma_1$ is the one-loop partial width, $\Gamma_0$
is the tree-level partial width and $\Delta^{loops}$ and $\Delta^{CT}$
provide the corrections coming from the one-loop diagrams and from the
counterterms, respectively.

Since the renormalization of the Higgs wave function, Higgs mass, the {\em{vevs}} (and hence $\tan\beta$) and the
para\-me\-ters $g$ and $M_W$ receive no ${\mathcal O}(\alpha_s)$
corrections at one-loop, the counterterms contribution originate just
from the renormalization of the quark wave functions and quark
masses. 
The corresponding expression for
$\Delta^{CT}$ is:
\be 
\Delta^{CT} = \frac{U_t}{D}\,\left(\frac{\delta m_t}{m_t}
        +\frac{1}{2}\,\delta Z_L^b+\frac{1}{2}\,\delta Z_R^t \,\right)+
  \frac{U_b}{D}\,\left( \frac{\delta m_b}{m_b}
        +\frac{1}{2}\,\delta Z_L^t+\frac{1}{2}\,\delta Z_R^b \, \right)\,,
\label{eq:CTdecay}
\ee 
where: 
\bea 
D &=&
(M_{H^+}^2-m_t^2-m_b^2)\,(m_t^2\cot^2\beta+m_b^2\tan^2\beta)
-4m_t^2m_b^2\,,\nonumber\\ 
U_t &=&(M_{H^+}^2-m_t^2-m_b^2)\,m_t^2\cot^2\beta - 2m_t^2m_b^2\,,\nonumber\\ 
U_b &=&(M_{H^+}^2-m_t^2-m_b^2)\,m_b^2\tan^2\beta - 2m_t^2m_b^2\,.
\eea 
We use the on-shell renormalization scheme as defined
in~\cite{Dabelstein}. The contributions from the counterterms can be
written in this scheme in terms of the bottom and
top self-energies as follows:
\bea
\frac{\delta m_{(t,b)}}{m_{(t,b)}}
        +\frac{1}{2}\,\delta Z_L^{(b,t)}+\frac{1}{2}\,\delta Z_R^{(t,b)} \,=
\Sigma^{(t,b)}_S(m_{(t,b)}^2)+ \frac{1}{2}\Sigma^{(t,b)}_L(m_{(t,b)}^2) -
\frac{1}{2}\Sigma^{(b,t)}_L(m_{(b,t)}^2)
\nonumber\\ 
- \frac{m_{t}^2}{2}
\left[\Sigma^{t'}_L(m_{t}^2) + \Sigma^{t'}_R(m_{t}^2)+ 2 \Sigma^{t'}_S(m_{t}^2)
\right] - \frac{m_b^2}{2}
\left[\Sigma^{b'}_L(m_b^2) + \Sigma^{b'}_R(m_b^2)+ 2 \Sigma^{b'}_S(m_b^2)
\right]\nonumber\\ 
\label{eq:CTselfE}
\eea 

\begin{figure}
\begin{center}
\epsfig{file=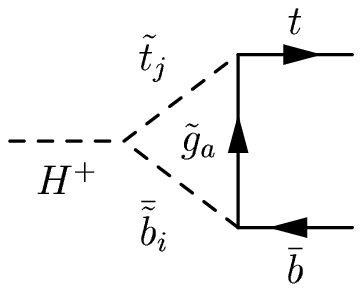,width=5cm}\vspace*{-0.7cm}\\
\epsfig{file=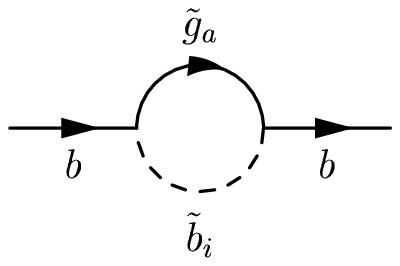,width=5cm}~~
\epsfig{file=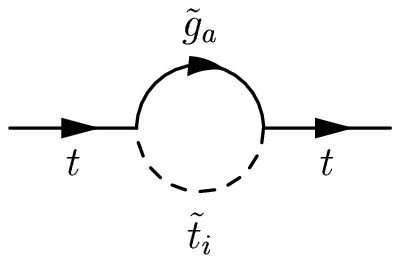,width=5cm}\vspace*{-0.3cm}
\caption{\it Diagrams corresponding to the vertex correction and quark
self-energies that contribute to $\Delta_{SQCD}$.}
\label{fig:diagramas}
\end{center}
\end{figure}

As we have already mentioned, in the present work we focus on the contributions to $\Delta^{loops}$ and $\Delta^{CT}$
coming from the SUSY-QCD sector:
\begin{equation}
\Delta_{SQCD} = \Delta_{SQCD}^{loops} + \Delta_{SQCD}^{CT}
\label{eq:defSQCD}
\end{equation}
where we use the short notation SQCD for SUSY-QCD. 

The SUSY-QCD contributions to the $H^+ \rightarrow t \bar b$ decay width
come from diagrams involving the exchange of virtual gluinos ($\tilde
g$), sbottoms ($\tilde b$) and stops ($\tilde t$). The diagrams that
contribute to the vertex corrections and quark self energies are shown
in Fig.~\ref{fig:diagramas}.
The contribution from the one-loop vertex diagrams are given by:
\be 
\Delta_{SQCD}^{loops} = \frac{U_t}{D}\,H_t+
\frac{U_b}{D}\,H_b\,,
\label{eq:loopsdecay}
\ee 
where:
\bea
 H_t=- \frac{2\alpha _s}{3 \pi}\,\frac{G_{ab}^*}{m_t
\cot\beta} [m_t R^{(t)}_{1 b}R^{(b)*}_{1 a}(C_{11}-C_{12})+ m_b R^{(t)}_{2
b}R^{(b)*}_{2 a}C_{12} \nonumber\\
+ M_{\tilde g} R^{(t)}_{2 b}R^{(b)*}_{1 a}C_0]
(m_t^2,M_{H^+}^2,m_b^2,M_{\tilde g}^2,M_{\tilde t_b}^2,M_{\tilde b_a}^2)\,,\nonumber\\
H_b=- \frac{2\alpha _s}{3 \pi}\, \frac{G_{ab}^*}{m_b \tan\beta} [m_t R^{(t)}_{2
b}R^{(b)*}_{2 a}(C_{11}-C_{12})+ m_b R^{(t)}_{1 b}R^{(b)*}_{1 a}C_{12}
\nonumber\\
+ M_{\tilde g} R^{(t)}_{1 b}R^{(b)*}_{2 a}C_0] (m_t^2,M_{H^+}^2,m_b^2,M_{\tilde g}^2,M_{\tilde t_b}^2,M_{\tilde b_a}^2)\,,
\label{eq:HLHR}
\eea
$R^{(q)}$ are given in eq.(\ref{eq:matrixrotation}) and we have used $G_{ab}$ to parametrize the $H^{+}\,\tilde
{b_a}\,\tilde{t_b}$ coupling. The expression for  $G_{ab}$ is given in
the Appendix A.

Finally, the contributions from the SUSY-QCD sector to
the quark self-energies are:
\bea
\Sigma^q_L(p^2)&=& - \frac{2\alpha_s }{3 \pi} 
\,|R^{(q)}_{1 a}|^2 B_1 (p^2,M^2_{\tilde{g}},M^2_{\tilde{q}_a})\,,\nonumber\\
\Sigma^q_R(p^2)&=& - \frac{2\alpha_s }{3 \pi}
\,|R^{(q)}_{2 a}|^2 B_1 (p^2,M^2_{\tilde{g}},M^2_{\tilde{q}_a})\,,\nonumber\\
\Sigma^q_S(p^2)&=& - \frac{2\alpha_s }{3 \pi}
\frac{m_{\tilde{g}}}{m_q}\, {\rm Re}(\,R^{(q)}_{1 a}\,R^{(q)*}_{2 a}) B_0(p^2,M^2_{\tilde{g}},M^2_{\tilde{q}_a})\,.
\label{eq:selfs}
\eea
We have decomposed the quark self-energy according to 
\begin{equation}
\Sigma^f(p)=\Sigma^f_L(p^2)\not{p}\,P_L+\Sigma^f_R(p^2)\not{p}\,P_R
+m_f\,\Sigma^f_S(p^2)\,,
\label{eq:Sigma}
\end{equation}
and used the notation $\Sigma'(p)\equiv \partial\Sigma(p)/\partial p^2$.

Our notation for the one-loop integrals that appear in the above formulae,
$B_0$, $B_0^{\prime}$, $B_1$, $B_1^{\prime}$,
$C_0$, $C_{11}$ and $C_{12}$, is defined in the Appendix B.
We have checked that these results are in agreement with the calculation
of~\cite{SQCD,QCDandS}.

\section{Large SUSY mass limit in the MSSM}
\label{sec:LargeMP}

Here we study the effect of
heavy SUSY particles in the SUSY-QCD correction to 
$H^+ \rightarrow t \bar b$. We derive approximate expansions
for this correction in the limit of large SUSY mass parameters as
compared to the electroweak scale, $M_{EW}$, which
corresponds to a SUSY spectrum heavier than the Standard Model
one. 

To define our expansion parameters, we consider all
the soft-SUSY-brea\-king mass parameters and the $\mu$ parameter to be
of the same order (co\-llec\-tively denoted by $M_{SUSY}$) and much
heavier than the electroweak scale.
That is,
\begin{equation}
    M_{SUSY} \sim M_{\tilde Q} \sim M_{\tilde U} \sim M_{\tilde D} 
\sim M_{\tilde g} \sim \mu \sim A_t \sim A_b \gg M_{EW},
\label{eq:largeSUSYmass}
\end{equation} 
where all these parameters are defined in
Section~\ref{sec:masses}. Notice that, with $M_{SUSY} \sim {\mathcal
O}(1 TeV)$, this choice will lead to a
plausible situation where all the SUSY particles in the SUSY-QCD sector
are much heavier than their SM partners.\footnote{Notice that our
choice of large $\mu$ is not a necessary condition to get all the
SUSY-QCD particles heavy but it is needed in order to get all the gauginos
heavy in the electroweak sector.}

We compute the expansions of the SUSY-QCD
correction to the partial decay width
$\Gamma(H^+ \rightarrow t \bar b)$ in inverse powers of the SUSY mass
parameters, up to order $M^2_{EW}/M^2_{SUSY}$, considering that all 
$M_{H^+}$, $M_Z$,
$M_W$, $m_t$ and $m_b$ are of order $M_{EW}$. In this section we
present the leading terms of
these expansions while the complete expressions are shown in the
Appendix C.

We consider here two possible extreme configurations of the squarks
mass-squared  matrices corresponding to maximal and
near-zero mixing respectively. This situation leads us to four extreme
cases that will be considered in the present work:
\begin{itemize}
\item Maximal mixing in the sbottom and stop sectors 
($\theta_{\tilde b} \sim \pm 45^o,\theta_{\tilde t} \sim \pm 45^o$)\,,
\item Near zero mixing in the sbottom and stop sectors ($\theta_{\tilde b}  \sim 0^o,\theta_{\tilde t} \sim  0^o$)\,,
\item Maximal mixing in the sbottom sector and near zero mixing in the
stop sector ($\theta_{\tilde b} \sim \pm 45^o,\theta_{\tilde t}  \sim 0^o$)\,,
\item Near zero mixing in the sbottom sector and maximal mixing in the
stop sector ($\theta_{\tilde b} \sim  0^o,\theta_{\tilde t}  \sim \pm 45^o$)\,.
\end{itemize}

Maximal mixing ($\theta_{\tilde q}\sim \pm 45^o$ ) arises when the
splitting between the dia\-go\-nal elements of the mass-squared matrix is
small compared to the off-diagonal elements, that is $|M_L^2 - M_R^2| \ll m_t
|X_t|$ in the stop sector or $|M_L^{'2} - M_R^{'2}| \ll m_b |X_b|$ in
the sbottom sector. This situation leads to small mass splitting
between the two squarks compared with the masses themselves, namely, 
$|M^2_{\tilde q_1} - M^2_{\tilde q_2}| \ll |M^2_{\tilde q_1} + M^2_{\tilde q_2}|$
. In order to define properly the large SUSY mass expansions in powers
of a small expansion parameter, we notice that all mass scales should
be referred either to $M_{EW}$ or to $M_{SUSY}$. Thus, the counting in
this maximal case goes as follows: $m_qX_q$ is of order
$M_{EW}M_{SUSY}$, while the splitting between the diagonal
mass-squared terms is of order $M^2_{EW}$, so that
the mass-squared eigenvalues can be written as:
\bea
M^2_{\tilde t_{1,2}} \simeq M_S^2 \pm \Delta_t^2\,\, \rm {or/and} &&\,\,\, 
M^2_{\tilde b_{1,2}} \simeq \tilde M_S^{2} \pm \Delta_b^2\,,
\label{eq:expMs}
\eea
where we have defined:
\bea
M_S^2 &=& \nicefrac{1}{2}(M_{\tilde t_1}^2 + M_{\tilde t_2}^2)  \,\,,\,\,\,  
\tilde M_S^2 = \nicefrac{1}{2}(M_{\tilde b_1}^2 + M_{\tilde b_2}^2)\,,
\label{eq:MsyMstilde}
\eea
\bea
 \Delta_t^2 &=& m_t |X_t| \left[1 + \frac{(M_L^2 -
M_R^2)^2}{8 m_t^2 X_t^2}\,\right]  \,, \nonumber \\
 \Delta_b^2 &=& m_b |X_b| \left[1 + \frac{(M_L^{'2} -
M_R^{'2})^2}{8 m_b^2 X_b^2}\,\right]\,,
\label{eq:MSandDeltat}
\eea
and $M_L$, $M_R$, $M^{'}_L$ and $M^{'}_R$ are defined in
eqs.~(\ref{eq:MLRtb}).
Here $M_S^2$ and $\tilde{M}_S^{2}$ are of order $M^2_{SUSY}$ while
$\Delta_t^2$ and $\Delta_b^2$ are of order $M_{EW}M_{SUSY}$. The
quantities $\frac{M^2_L - M^2_R}{m_tX_t}$ and 
$\frac{M^{'2}_L - M^{'2}_R}{m_bX_b}$ are therefore small and of order 
$M_{EW}/M_{SUSY}$. Expan\-ding the 
expressions for the mixing angle in terms of these small parameters, we 
obtain for $\theta_{\tilde q} \simeq 45^o$:
\bea
\label{eq:Maxmsct}
    \cos 2 \theta_{\tilde t} \simeq
    \frac{M_L^2 - M_R^2}{2 m_t X_t} \,\, &,& \,\,\,
    \cos 2 \theta_{\tilde b} \simeq
    \frac{M_L^{'2} - M_R^{'2}}{2 m_b X_b} 
     \,, \nn \\
    \sin 2 \theta_{\tilde t} \simeq
     \sigma_{X_t} &&\left[1 - \frac{(M_L^2 - M_R^2)^2}{8
    m_t^2 X_t^2} \right] \,,\nonumber\\
    \sin 2 \theta_{\tilde b} \simeq
     \sigma_{X_b} &&\left[1 - \frac{(M_L^{'2} -
    M_R^{'2})^2}{8 m_b^2 X_b^2} \right] \,.
\eea
where $\sigma_{X_q} \equiv sgn(X_q)$ with $q=t,b$.

Near-zero mixing ($\theta_{\tilde q} \simeq 0^o$) arises when the
splitting between the diagonal elements of the mass-squared matrix is
large compared to the off-diagonal elements, that is $|M_L^2 - M_R^2| \gg m_t
|X_t|$ in the stop sector or $|M_L^{'2} - M_R^{'2}| \gg m_b |X_b|$ in
the sbottom sector. This is the case usually
considered in the lite\-ra\-tu\-re, because $M_L$ ($M^{'}_L$) and $M_R$
($M^{'}_R$) depend on two different (and a priori independent) 
soft-SUSY-breaking parameters
$M_{\tilde Q}$ and $M_{\tilde U}$ ($M_{\tilde D}$), respectively. 
The counting in terms of $M_{EW}$ and $M_{SUSY}$ is such that 
$(M_L^{2} - M_R^{2})$ (or $(M_L^{'2} - M_R^{'2})$) is of order
$M^2_{SUSY}$ while $m_qX_q$ is of order $M_{EW}M_{SUSY}$. The
mass splitting between the two squarks will be of the same order as
the masses themselves, that is 
$|M^2_{\tilde q_1} - M^2_{\tilde q_2}| \sim {\mathcal O} (|M^2_{\tilde
q_1} + M^2_{\tilde q_2}|) \sim {\mathcal O}(M^2_{SUSY})$.
As in the maximal mixing case, all mass scales should be referred to
$M_{EW}$ or to $M_{SUSY}$ in order to define the expansion
parameters. In this case we write our results in terms of the physical stop and sbottom
masses and the mixing angles which are given by:
\begin{equation}
    \sin 2 \theta_{\tilde q} =
    \frac{2 m_q X_q}{M^2_{\tilde q_1} - M^2_{\tilde q_2}}\,,
\label{eq:zeromst}
\end{equation}
\begin{equation}
\label{eq:zeromct}
    \cos 2 \theta_{\tilde q} \simeq
    1 - \frac{2 m_q^2 X_q^2}{(M_{\tilde q_1}^2 - M_{\tilde q_2}^2)^2}\,.
\end{equation}

\subsection{The decoupling regime in $H^+ \rightarrow t \bar b$ decay}

Now we study, both analytically and numerically, the SUSY-QCD correction to $\Gamma (H^+ \rightarrow t \bar
b)$ in the large SUSY mass parameters limit. We consider here all the
possible extreme configurations of the squarks
mass-squared matrix, namely the four cases mentioned above. We will
study also the
individual decoupling of the various SUSY particles, that is, decoupling
of gluinos and squarks separately. In this analysis we use the leading term of
our expansions for
$\Delta_{SQCD}$ while the complete expressions, up to order
$M_{EW}^{2}/M_{SUSY}^{2}$, are given in the Appendix C.

First we consider maximal mixing in both the stop and sbottom
sectors. 
In this case, and in the large SUSY mass parameters
limit defined in eq.~(\ref{eq:largeSUSYmass}), we obtain $M_L^{2} \simeq
M_R^{2}$ and $M_L^{'2} \simeq M_R^{'2}$, up to corrections of order
$M_{EW}^{2}/M_{SUSY}^{2}$. Since $M_L$ and $M_L^{'}$ are determined by
the same soft-SUSY breaking
term $M_{\tilde Q}$, we also have $M_L^{2} \simeq M_L^{'2}$. From these
relations and from eq.~(\ref{eq:MSandDeltat}) we obtain $M_S^{2} \simeq \tilde M_S^{2}$, up to  corrections of order
$M_{EW}^{2}/M_{SUSY}^{2}$. 

Taking this into account we expand the SUSY-QCD contributions to the 
$H^{+} \rightarrow t \bar b$ partial width and we find:
\begin{equation}
\Delta_{SQCD}= - \frac{\alpha_s}{3 \pi} \frac{M_{\tilde g}
\mu}{M_S^{2}} (\tan \beta + \cot \beta ) f_1 (R) + \mathcal{O} 
(M_{EW}^{2}/M_{SUSY}^{2})
\label{eq:nondecbothmax}
\end{equation}
where $R=M_{\tilde g}/M_S$. The expression for $f_1 (R)$, which  arises
from the expansions of the loop integrals, is given in Appendix B and
it is normalized so that $f_1 (1)=1$.
The complete formula for $\Delta_{SQCD}$, including the ${\mathcal
O}(\frac{M^2_{EW}}{M^2_{SUSY}})$ terms, is given in Appendix C.

We can see from eq.~(\ref{eq:nondecbothmax}) that, taking all SUSY mass
parameters arbitra\-ri\-ly large and of the same order,
$\Delta_{SQCD}$ leads to a non-zero
value. That is, the SUSY-QCD correction {\it do not decouple} in the large
SUSY mass parameters regime. This can be seen clearly, for instance, in the
simplest case of equal mass scales, 
$\mu=M_{\tilde g}=M_S$, where $f_1(R)=1$, leading to
\begin{equation}
\Delta_{SQCD}= - \frac{\alpha_s}{3 \pi} (\tan \beta + \cot \beta )
+ \mathcal{O} (M_{EW}^{2}/M_{SUSY}^{2}),
\end{equation}
which shows that the first term in the large $M_{SUSY}$ expansion is
indeed of $\mathcal{O} (M_{EW}^{0}/M_{SUSY}^{0})$.
This leading contribution, 
when considered in the large $\tan \beta$ regime and 
expressed in terms of an effective coupling of $H^+$ to $b \bar t$ is in
agreement with the previous results of refs.~\cite{CarenaDavid} 
that were obtained in the zero external momentum 
approximation by using an effective Lagrangian approach. 

Notice also that this non-decoupling term is enhanced by a 
$\tan \beta$ factor and therefore a numerically large SUSY-QCD correction in the large 
$\tan \beta$ limit is expected.
\begin{figure}[tb]
\begin{center}
\epsfig{file=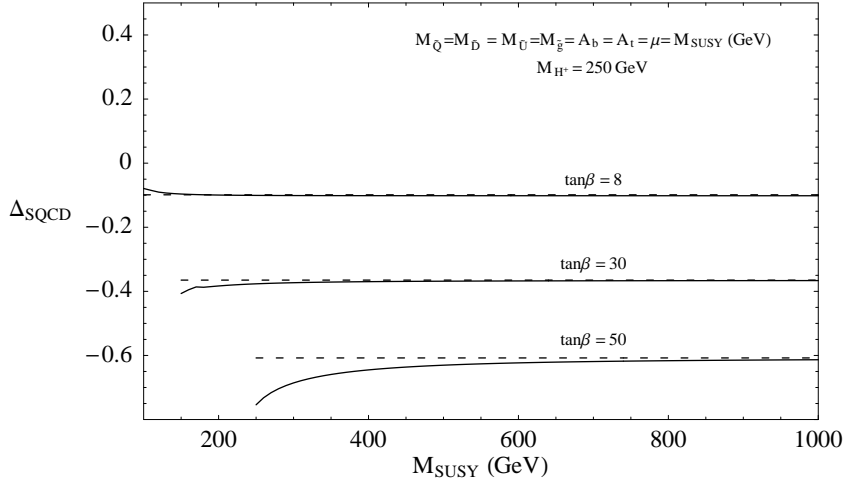,width=12cm}\vspace*{-0.3cm}
\caption{\it Non-decoupling behavior of $\Delta_{SQCD}$ with
$M_{\tilde Q} = M_{\tilde U} = M_{\tilde D} = M_{\tilde g} = A_b = A_t
= \mu = M_{SUSY}$ and for different values of $\tan
\beta$. Exact one-loop result (solid
lines) and the expansion given in  eq. (\ref{eq:nondecbothmax}) (dashed
lines) are plotted for comparison.}
\label{fig:Unaescala}
\end{center}
\end{figure}
This non-decoupling behavior is shown numerically in
Fig.~\ref{fig:Unaescala}. 
In this figure we plot the exact one-loop result (solid
lines) and the leading term of the large SUSY mass expansion in 
eq.~(\ref{eq:nondecbothmax}) (dashed
lines) for $M_{\tilde Q} = M_{\tilde U} = M_{\tilde D} = M_{\tilde
g} = A_b = A_t = \mu = M_{SUSY}$ and for different values of $\tan
\beta$. For all the numerical analyses, in this work we fix $M_{H^+} =
250$ GeV. 
We can see in this figure that for large SUSY mass parameters (say
$M_{SUSY} \geq 300$ GeV), $\Delta_{SQCD}$
tends to a non-zero
value whose size is larger for larger $\tan \beta$
values. Indeed the correction can be quite large for large $\tan
\beta$ even for quite heavy SUSY particles. For example for $M_{SUSY} =
1$ TeV and $\tan \beta = 30$ we get $\Delta_{SQCD} \simeq - 35 \%$, and
it grows linearly with $\tan \beta$.

From the numerical comparison in Fig.~\ref{fig:Unaescala} between the exact and
our approximate result of eq. (\ref{eq:nondecbothmax}), we can conclude that
this expansion, with just the leading
term, is a good approximation for
large enough SUSY mass para\-me\-ters. It is clear also that as $\tan
\beta$ grows the agreement between the exact and approximate results
becomes worse at low $M_{SUSY}$. However, we can conclude that the
agreement is still quite good even for $\tan \beta$ as large as 50
whenever $M_{SUSY} \geq 300$ GeV.

\begin{figure}[tb]
\begin{center}
\epsfig{file=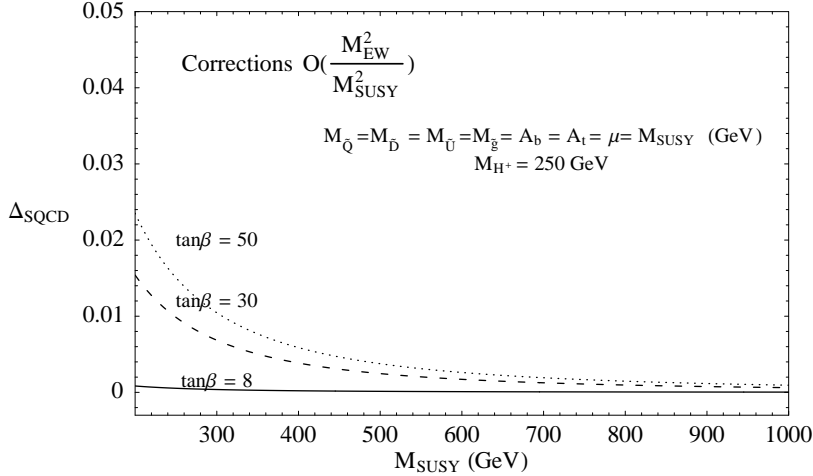,width=12cm}\vspace*{-0.5cm}
\caption{\it Behaviour of the contributions to $\Delta_{SQCD}$ of $\mathcal{O} 
(M_{EW}^{2}/M_{SUSY}^{2})$ for
$M_{\tilde Q} = M_{\tilde U} = M_{\tilde D} = M_{\tilde g} = A_b = A_t
= \mu = M_{SUSY}$ and for different values of $\tan
\beta$.}
\label{fig:anal}
\end{center}
\end{figure}

For lower values of $M_{SUSY}$ the next to leading co\-rrec\-tions,
i.e. the co\-rrec\-tions of order $M_{EW}^2/M_{SUSY}^2$ which are given in
Appendix C, begin to be important. We have estimated the size of these
corrections for several choices of the parameter space and we have
found that they are of order of a few percent. Besides, they
grow with the $\mu$ parameter and with $\tan \beta$ and are
almost independent of the value for the trilinear couplings,
$A_{t,b}$. We show in Fig.~\ref{fig:anal} the numerical results for
the particular case $M_{\tilde Q}= M_{\tilde U} = M_{\tilde D} = M_{\tilde g} 
= A_b = A_t = \mu = M_{SUSY}$ and for several values of $\tan \beta$. We
can see clearly that their size is always small as compared to the
leading non-decoupling term. For other choices of the parameters
similar results are found.

In eq.~(\ref{eq:nondecbothmax}) we can see that, in the approximation
of large SUSY mass parameters, chan\-ging
the sign of $M_{\tilde g} \mu$ simply flips the sign of the 
SUSY-QCD correction. In most of this work, and for the numerical analysis, 
a positive sign for $M_{\tilde g} \mu$ has been chosen.

\begin{figure}[t]
\hspace{-1.7cm}\epsfig{file=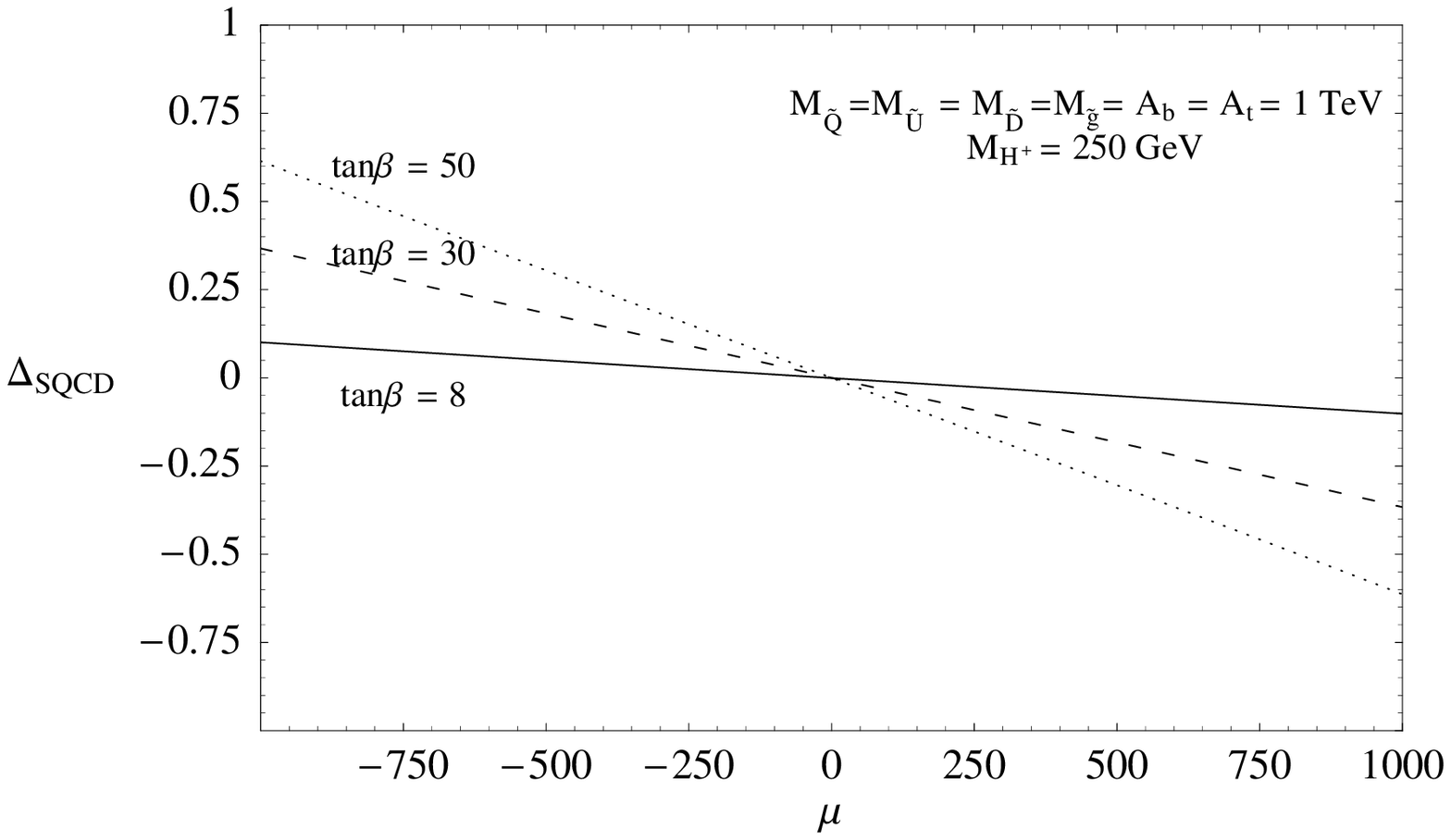,height=6cm,width=8cm}~~
\epsfig{file=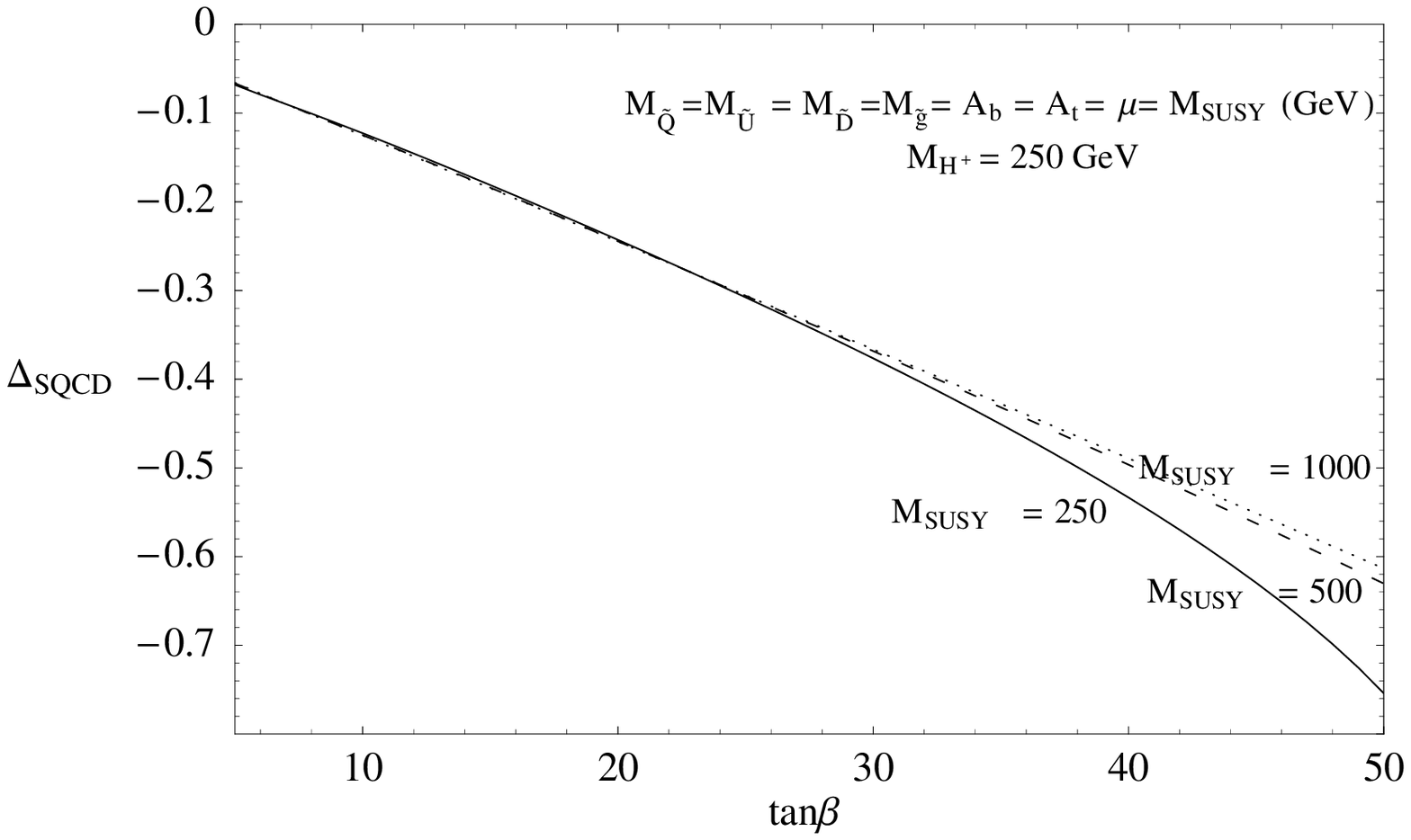,height=6cm,width=8cm}\vspace*{-0.7cm}
\caption{\it $\Delta_{SQCD}$ as a function of $\mu$ and
$\tan \beta$.}
\label{fig:Depmuandtb}
\end{figure}

In Fig.~\ref{fig:Depmuandtb} we show the exact SUSY-QCD correction as a function of
$\mu$ (left) and as a function of $\tan \beta$ (right). In the left
panel we plot $\Delta_{SQCD}$ for differents values of $\tan \beta$
and fixing $M_{\tilde Q}= M_{\tilde U} = M_{\tilde D} = M_{\tilde g} 
= A_b = A_t = 1$ TeV. We see that the size of the correction grows linearly
with $\mu$ and notice that by changing the sign of $\mu$ the sign of
the correction changes.
In the right panel we plot the SUSY-QCD correction as a function
of $\tan \beta$ for the choice  
$M_{\tilde Q} = M_{\tilde U} = M_{\tilde D} = M_{\tilde
g} = A_b = A_t = \mu = M_{SUSY}$ and for different values of
$M_{SUSY}$.  We see explicitely that for large $M_{SUSY}$ the size of
the correction grows linearly with $\tan \beta$.

\begin{figure}[tb]
\hspace{-1.7cm}\epsfig{file=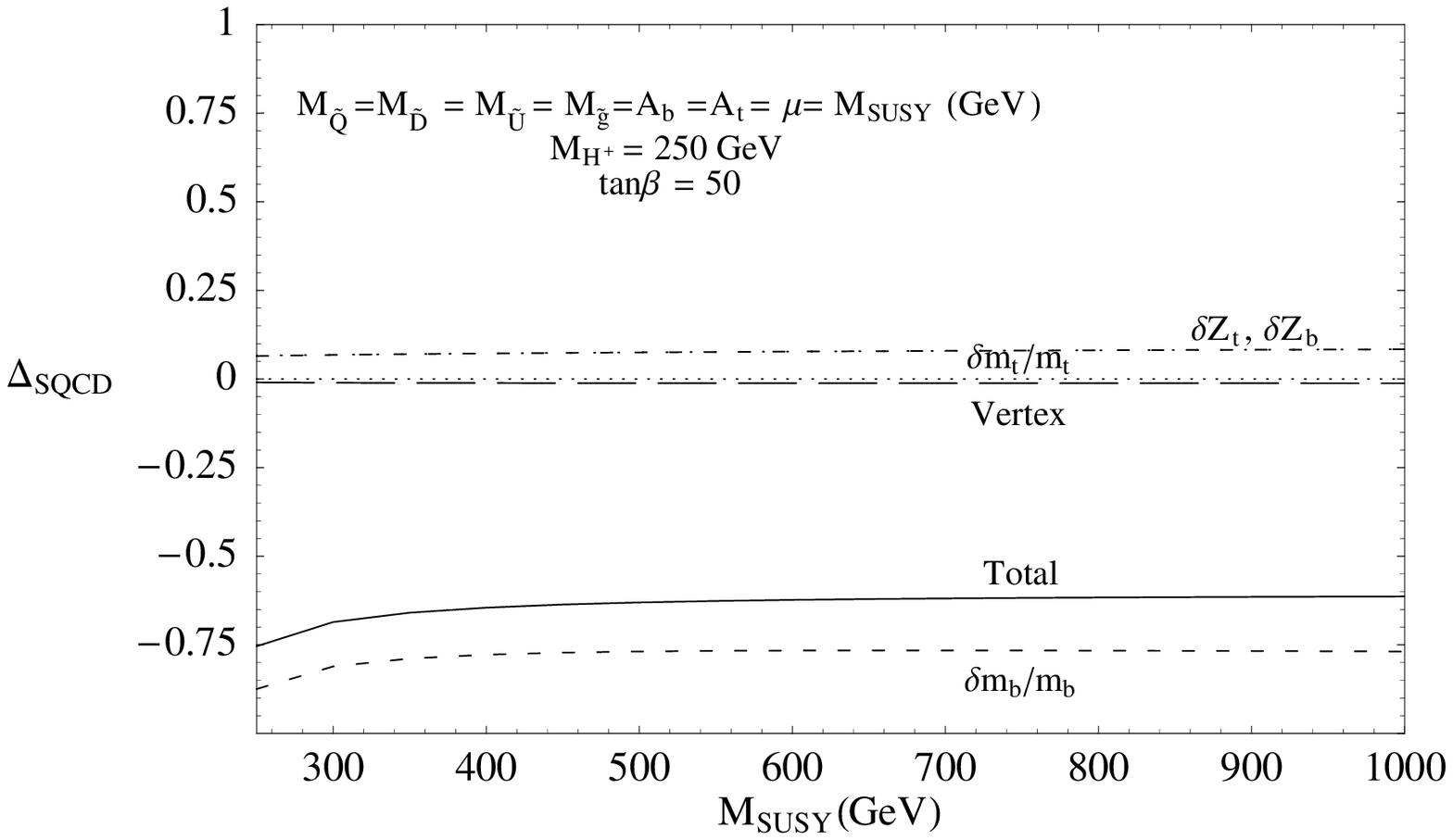,height=6cm,width=8cm}~~
\epsfig{file=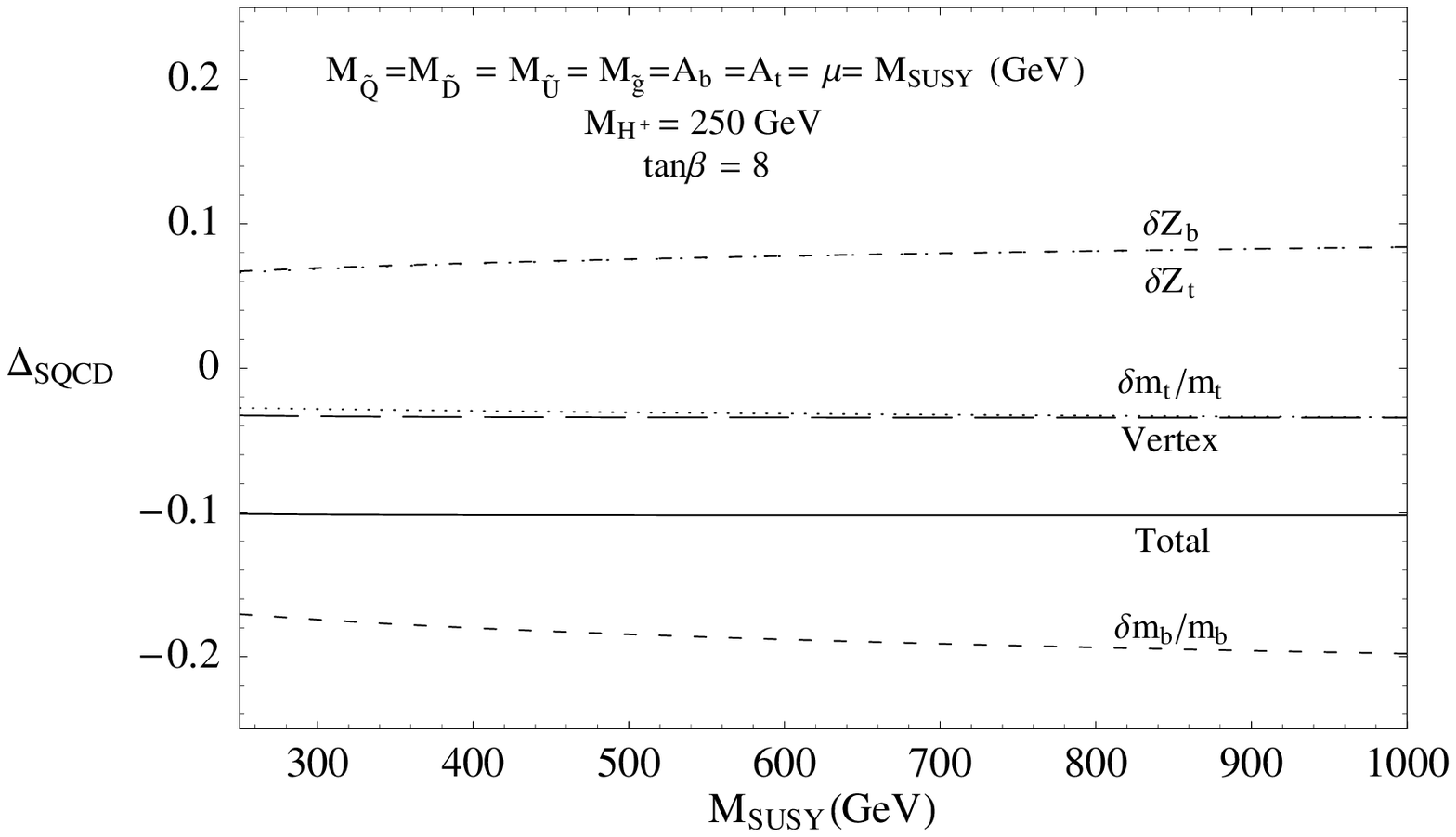,height=6cm,width=8cm}\vspace*{-0.3cm}
\caption{\it Comparison of contribution coming from $\delta m_b/m_b$
(lower dashed line), $\delta Z_b$ (upper dashed line), $\delta
m_t/m_t$ ( lower dotted line), $\delta Z_t$ (upper dotted line) and vertex
contributions (long-dashed line) for different values
of $\tan \beta$. Total correction is also plotted (solid line).}
\label{fig:mtandmbsep}
\end{figure}

In comparing the various contributions to $\Delta_{SQCD}$ we find that
the dominant contributions come from the renormalization of the
bottom-quark 
mass and wavefunction. In Fig.~\ref{fig:mtandmbsep} the total SUSY-QCD
correction (Total), the contributions from the bottom-quark mass 
($\delta m_b/m_b$) and wavefunction ($\delta Z_b$) counterterms, the 
contributions from the top-quark mass 
($\delta m_t/m_t$) and wavefunction ($\delta Z_t$) counterterms and
the vertex contributions (Vertex) are shown separately. We can see
that the contributions from $\delta m_b/m_b+\delta Z_b$ dominate 
$\Delta_{SQCD}$ at large $\tan \beta$.

For the three cases left, near-zero stop and sbottom mixing, near-zero
stop-maximal sbottom mixing and near-zero sbottom-maximal stop mixing,
we have found a leading term in $\Delta_{SQCD}$ that is very
similar to the one in 
eq.~(\ref{eq:nondecbothmax}). The general expression is:
\begin{equation}
\Delta_{SQCD}= - \frac{\alpha_s}{3 \pi} \frac{M_{\tilde g}
\mu}{\hat M_S^{2}} (\tan \beta + \cot \beta ) F_{mix} + \mathcal{O} 
(M_{EW}^{2}/M_{SUSY}^{2})\,,
\label{eq:nondecgen}
\end{equation}
where $\hat M_S^{2}$ and $F_{mix}$ depend on the case:
\begin{itemize}
\item Near-zero stop and sbottom mixing:  

$\hat M_S^{2} = M_{\tilde t_1}^{2}$ and $F_{mix}= f_1 (R_1,R_2)$ where
$R_i=M_{\tilde g}/M_{\tilde q_i}$ and 

$M_{\tilde t_i} \simeq M_{\tilde
b_i} \simeq M_{\tilde q_i}$, i=1,2 \,.
\item Near-zero stop-maximal sbottom mixing:

$\hat M_S^{2} = \tilde M_{S}^{2}$ and $F_{mix}= \frac{U_t}{D}f_1
(R_b,R_{t_2}) +\frac{U_b}{D}f_1 (R_b) $ where 
$R_{t_2}=M_{\tilde g}/M_{\tilde t_2}$ and $R_b=M_{\tilde g}/\tilde M_{S}$.
\item Near-zero sbottom-maximal stop mixing:

$\hat M_S^{2} = M_{S}^{2}$ and $F_{mix}= \frac{U_t}{D}f_1
(R_t) +\frac{U_b}{D}f_1 (R_t,R_{b_2}) $ where 
$R_{b_2}=M_{\tilde g}/M_{\tilde b_2}$ and $R_t=M_{\tilde g}/M_{S}$.
\end{itemize}
The functions $f_1(R)$ and $f_1(R_1,R_2)$ are defined in the Appendix
B and are normalized
such that $f_1(1,1)= f_1(1)=1$. $M_S$ and $\tilde M_S$ are defined in
eq. (\ref{eq:MsyMstilde}). The corrections in eq. (\ref{eq:nondecgen}) of
${\mathcal O}(\frac{M^2_{EW}}{M^2_{SUSY}})$ are given explicitely in
Appendix C.

In eq.~(\ref{eq:nondecgen}) we see that the non-decoupling
contribution to $\Delta_{SQCD}$ appears in all the cases so that 
we can conclude that 
the SUSY-QCD corrections {\it do not decouple} in the large
SUSY mass parameters regime. We also see that $\Delta_{SQCD}$ grows
again linearly with $\tan \beta$, thus we expect large corrections in the 
large $\tan \beta$ limit for all the cases.

Now we study the decoupling behaviour of the SUSY-QCD corrections as indivi\-dual
SUSY particles become heavier than a reference SUSY mass scale so that
there is a hierarchy among the various SUSY particle masses. We
will consider two cases: large gluino mass with maximal mixing in stop
an sbottom sectors and large squark masses (stop and sbottom) with
maximal mixing in both sectors.
\begin{figure}[t]
\begin{center}
\epsfig{file=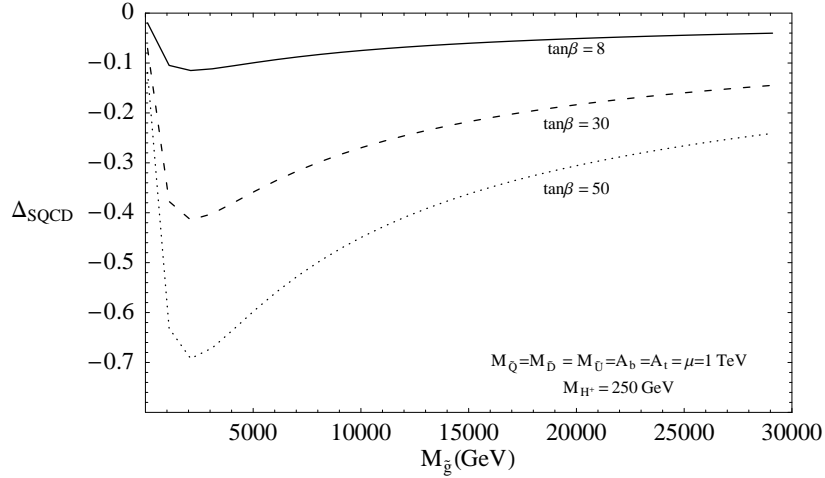,width=12cm}\vspace*{-0.6cm}
\caption{\it Behaviour of $\Delta_{SQCD}$ in the large $M_{\tilde g}$
limit with fixed $M_{\tilde Q} = M_{\tilde
U} = M_{\tilde D} = A_b = A_t = \mu = 1$ TeV and for different values
of $\tan \beta$.}
\label{fig:DepMg}
\end{center}
\end{figure}

First, we examine the case of a very heavy gluino compared with the rest
of the SUSY spectrum and with maximal mixing in
sbottom and stop sectors. In this case $M_{\tilde g} \gg M_{SUSY} \gg M_{EW}$. We evaluate the SUSY-QCD
correction when the gluino mass is taken very large while the rest of
the SUSY mass parameters remain fixed to a reference value of the
order of $M_{SUSY}$ (larger than the electroweak
scale). Our result for $\Delta_{SQCD}$ is:
\begin{equation}
\Delta_{SQCD}= \frac{2 \alpha_s}{3 \pi} (\tan \beta + \cot \beta
) \left( 1 - \log \frac{M_{\tilde g}^{2}}{M_S^{2}}\right) \frac{\mu}{M_{\tilde g}}+ \mathcal{O} 
(\frac{M_{EW}^{2}}{M_{SUSY}M_{\tilde g}})\,.
\label{eq:gluinodec}
\end{equation}

In eq.~(\ref{eq:gluinodec}) we see that the SUSY-QCD correction {\it
decouples} when we take the large gluino mass limit while the rest of
the SUSY mass parameters ($\mu$ and $M_{S}$) remain fixed at some
scale of the order $M_{SUSY}$. We also notice that this decoupling is
very slow: $\Delta_{SQCD}$
falls off when increasing $M_{\tilde g}$ as $\frac{\log M_{\tilde
g}}{M_{\tilde g}}$. We see again that the correction is
enhanced by a $\tan \beta$ factor, so we expect large SUSY-QCD corrections
in the large $\tan \beta$ limit. In Fig.~\ref{fig:DepMg} we plot the
exact result for
$\Delta_{SQCD}$ as a function of $M_{\tilde g}$ and keeping the other
SUSY mass parameters fixed at 1 TeV. We can see the slow decoupling with
the gluino mass and the numerically large correction even for very
heavy gluino; for example if $M_{\tilde g}= 2$ TeV and $\tan \beta =
30$ we find $\Delta_{SQCD} \simeq -40 \%$. Notice that the size can be 
so large that the validity of the perturbative expansion can be
questionable. We refer the reader to refs.~\cite{CarenaDavid} 
where this subject is studied in more detail and some techniques 
of resummation 
for a better convergence of the series are proposed.

Next we analyze the large squark (stop and sbottom) mass limit with
maximal mixing in both sectors. We consider the case with $M_{S} = \tilde
M_{S} \gg M_{\tilde g} = A_t = A_b = \mu \gg M_{EW}$. If we expand the
eq.~(\ref{eq:nondecbothmax}) in the limit of large squark masses we
obtain:
\begin{equation}
\Delta_{SQCD}= - \frac{2 \alpha_s}{3 \pi} \frac{M_{\tilde g}
\mu}{M_S^{2}} (\tan \beta + \cot \beta ) + \mathcal{O} 
(M_{EW}^{2}/M_{SUSY}^{2})\,.
\label{eq:squarkdec}
\end{equation}
\begin{figure}[t]
\begin{center}
\epsfig{file=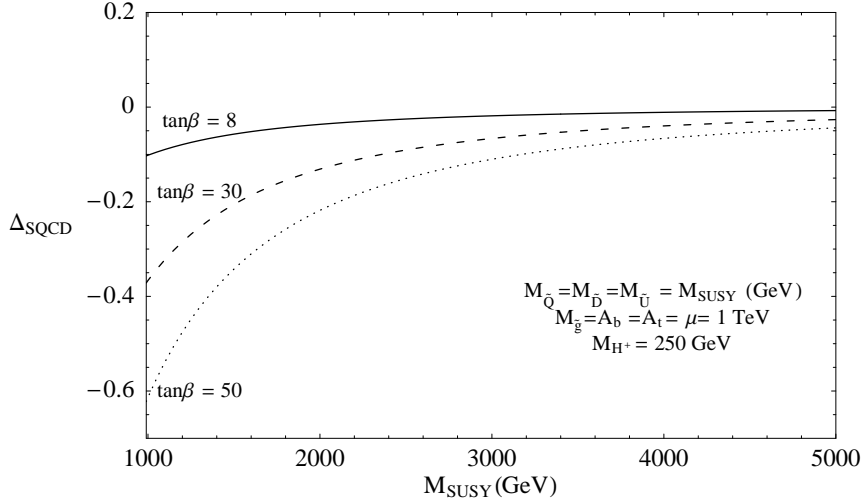,width=12cm}\vspace*{-0.6cm}
\caption{\it Behaviour of $\Delta_{SQCD}$ in the large squark masses
limit with fixed $M_{\tilde g} = A_b = A_t = \mu
= 1$ TeV and for different values of $\tan \beta$.}
\label{fig:DecMsq1TeV}
\end{center}
\end{figure}
As we can see in  eq.~(\ref{eq:squarkdec}), the correction {\it
decouples} in the large squark mass limit if the rest of the SUSY mass
parameters ($M_{\tilde g}$ and $\mu$) remain fixed at some scale
larger than the electroweak scale. This decoupling is faster than the
previous case
since $\Delta_{SQCD}$ is proportional to the inverse of $M_{S}^{2}$,
that is the scale of the squark masses. We also see that the
correction is enhanced by a $\tan \beta$ factor. Finally we have also
confirmed this decoupling behaviour of squarks numerically, as can be
seen in
Fig.~\ref{fig:DecMsq1TeV}, where the explicit result of $\Delta_{SQCD}$,
as a function of $M_{SUSY}=M_{\tilde Q}=M_{\tilde D}=M_{\tilde U}$, is
shown for several values of $\tan \beta$.

\section{Conclusions}
\label{sec:conclusions}

In this work we have looked for indirect SUSY signals through the
effect of radiative corrections to the $H^+ \rightarrow t\bar b$
decay in the limit of large SUSY masses. We have focused on the well
known large SUSY-QCD corrections to the $\Gamma (H^+ \rightarrow t\bar b)$
partial decay width. We have studied in detail
the one-loop SUSY-QCD corrections to $\Gamma (H^+ \rightarrow t\bar b)$
when the SUSY particles are heavier than the SM ones. In order
to understand analytically the behaviour of these corrections in the
large SUSY mass limit, we have performed expansions of the one-loop
partial width that
are valid for large values of the SUSY mass parameters compared to the
electroweak scale and for various choices of the mixing in the squarks
sector. 

We have shown that for large SUSY mass parameters and all of the same
order, the SUSY-QCD radiative corrections do not decouple in the $H^+
\rightarrow t\bar b$ decay. In other words, if there are heavy squarks
and gluinos, with masses of the same order and much larger than the SM
particles, they will
produce a non-vanishing contribution, via SUSY-QCD radiative
corrections, to the potentially measurable partial width $\Gamma (H^+
\rightarrow t\bar b)$.
We would like to remind that the conclusion about decoupling or
non-decoupling of SUSY particles in the MSSM is not inmediate, since
the decoupling theorem  does not apply to this case, because the MSSM
is a theory involving chiral fermions and the spontaneous brea\-king of the
electroweak symmetry. Therefore, the SUSY decoupling behaviour must be
explored case by case. In this direction it has been shown by an
explicit calculation~\cite{TesisS} that there
is decoupling of heavy SUSY particles in observables with external
electroweak gauge bosons.

We interpret the {\it non-decoupling} effect found in the present work (as
well as the one found in~\cite{HaberTemes}) as follows. If we start
from the full MSSM as the theory valid at large energies and integrate
out the heavy SUSY spectrum, we are left with a low energy theory,
valid at the electroweak scale, which contains the Standard Model
particles plus two full Higgs doublets. The interactions of this
effective low energy theory are a priori unknown, but they can be
derived by explicit integration of the heavy SUSY
particles~\cite{dht}. The Higgs-fermion-fermion interactions of the
MSSM are like those of a two Higgs doublet model (2HDM) of type II.
In a 2HDM of type II there are some restrictions on the allowed
couplings, which in the MSSM case are a consequence of
supersymmetry. Once the heavy SUSY particles are integrated out, new
low energy effective Higgs-fermion-fermion interactions with no
restrictions anymore can be generated since SUSY is a (softly)
broken symmetry. Indeed, one expects to find the most general 2HDM of
type III, where both Higgs doublets can couple to both {\it u} and
{\it d} type fermions. However, to show this one must perform
explicitly the computation and find out the specific values of the
generated effective couplings~\cite{dht}.

We also emphasize here that these non-decoupling SUSY-QCD corrections
could give us an indirect signal of supersymmetry at present and future
colliders even for a very heavy SUSY spectrum at the TeV scale. In particular they can provide some clues in the indirect
search of a heavy SUSY spectrum in the LHC~\cite{chtt} and next linear
colliders.

We have also examined in this work some special cases in which there is a hierarchy among
the SUSY mass parameters.  In the case of maximal squark mixing with
$M_S=\tilde M_S$ large and the other SUSY mass parameters of order a
common mass scale
$M$ (chosen such that $M_{EW}\ll M\ll M_S$), the SUSY-QCD
corrections decouple like $M^2/M_S^2$.
In addition, we have examined the case of a large gluino mass with the other SUSY
mass parameters of order a common mass scale $M$
(chosen such that $M_{EW}\ll M\ll M_{\tilde g}$).
In this case we have found that the SUSY-QCD corrections
decouple more slowly, like
$(M/M_{\tilde g}) \log(M^2_{\tilde g}/M_S^2)$. This indicates that
sizeable indirect signals from gluinos as heavy as several TeV can be
obtained in the $H^+ \rightarrow t\bar b$ decay.

Finally we have seen that, in all cases, the corrections are enhanced
by a $\tan \beta$ factor and therefore large SUSY-QCD contributions to the
$H^+ \rightarrow t\bar b$ decay are expected for large values of $\tan
\beta$, even for a very heavy SUSY spectrum.


\section*{Acknowledgments}
This work has been supported in part by
the Spanish Ministerio de Ciencia y Tecnolog{\'\i}a under project CICYT
FPA 2000-0980. S.P. was partially su\-ppor\-ted by the
Ramon Areces Foundation at the Karlsruhe University.
S.P. wishes to thank W.~Hollik for reading the manuscript.


\section*{Appendices}
\vspace{0.4cm}
\setcounter{equation}{0}

\section*{{\bf{A}} $H^{\pm}\,\tilde t\,\tilde{b}$  interaction Lagrangian}
\renewcommand{\theequation}{A.\arabic{equation}}

The $H^{+}\,\tilde t^{*}\,\tilde{b}$ and $H^{-}\,\tilde t\,\tilde{b}^{*}$
interaction terms are given by the following interaction Lagrangian~\cite{MSSMHiggssector}:
\be
\mathcal{L}_{H^{\pm}\,\tilde t\,\tilde{b}} = - \frac{g}{\sqrt{2}M_W}
H^{-} \left( g_{LL} \, \tilde b_L^{*} \tilde t_L +  g_{RR} \, \tilde b_R^{*}
\tilde t_R +  g_{LR} \, \tilde b_R^{*} \tilde t_L +  g_{RL} \, \tilde b_L^{*}
\tilde t_R
\right) + h.c. \,,
\ee
where:
\bea
g_{LL}&=& M_W^2 \sin 2\beta - (m_t^2 \cot \beta + m_b^2 \tan \beta)\,,
\nonumber \\
g_{RR}&=& - m_t m_b (\tan \beta + \cot \beta)\,, \nonumber \\
g_{LR}&=& - m_b (\mu + A_b \tan \beta) \,, \nonumber \\
g_{RL}& =& - m_t (\mu + A_t \cot \beta) \,. 
\eea
In the mass eigenstate basis for the squarks, these interaction terms
can be parametrized as follows:
\be
\mathcal{L}_{H^{\pm}\,\tilde t\,\tilde{b}} = - \frac{g}{\sqrt{2}M_W}
H^{-} G_{ab} \, \tilde b_a^* \tilde t_b + h.c.
\ee
where
\be 
G_{ab}=R^{(b)*}_{1 a}R^{(t)}_{1 b}  
g_{LL}+ R^{(b)*}_{2 a}R^{(t)}_{2 b} g_{RR}+ R^{(b)*}_{2 a}R^{(t)}_{1
b} g_{LR}+
R^{(b)*}_{1 a}R^{(t)}_{2 b} g_{RL}\,. 
\ee 

\section*{{\bf{B}} Large mass expansion of loop integrals}
\renewcommand{\theequation}{B.\arabic{equation}}

Here we define the notation for the two- and three-point
integrals that appear in eqs.~(\ref{eq:HLHR}) and (\ref{eq:selfs})
and give formulae for their expansions in inverse powers of the SUSY mass
parameters.

We follow the definitions and conventions of~\cite{Hollik}. The
integrals are performed in $D=4 - \epsilon$ dimensions and the divergent
contributions are regularized by $\Delta = \frac{2}{\epsilon} - \gamma_E +
\log(4\pi)$, with the corresponding inclusion of the energy scale
given here by $\mu_0$.

The two-point integrals are given by:
\begin{equation}
    \mu_0^{4-D} \int \frac{d^D k}{(2\pi)^D} \frac{\{1; k^{\mu}\}}
    {[k^2 - m_1^2][(k+q)^2 - m_2^2]}
    = \frac{i}{16\pi^2} \left\{ B_0; q^{\mu} B_1 \right\}
    (q^2;m_1^2,m_2^2)\,.
\end{equation}
The derivatives of the two-point functions are defined as follows:
\begin{equation}
    B^{\prime}_{0,1}(p^2;m_1^2,m_2^2)
    = \left. \frac{\partial}{\partial q^2} B_{0,1}(q^2;m_1^2,m_2^2)
    \right|_{q^2 = p^2}\,.
\end{equation}
Finally, the three-point integrals are given by:
\bea
    &\mu_0^{4-D}& \int \frac{d^D k}{(2\pi)^D}
    \frac{\{1; k^{\mu}\}}
    {[k^2 - m_1^2][(k + p_1)^2 - m_2^2][(k + p_1 + p_2)^2 - m_3^2]}
    \nonumber \\
    & & = \frac{i}{16\pi^2}
    \left\{ C_0; p_1^{\mu}C_{11} + p_2^{\mu}C_{12} \right\}
    (p_1^2, p_2^2, p^2; m_1^2, m_2^2, m_3^2)\,,
\eea

\vspace{-0.1in}\noindent
where $p = -p_1 - p_2$.

We next give the large $M_{SUSY}$ expansions of the one-loop integrals in the
four different cases that we have mentioned in section~\ref{sec:LargeMP}.

\subsection*{{\bf{B.1}} Maximal sbottom and stop mixing}

The results of the large $M_{SUSY}$ expansions of the loop integrals
are as follows: 
\bea
\label{eq:maxbt}
&& C_0(m_t^2,M_{H^+}^2,m_b^2;M_{\tilde g}^2, M_{\tilde t_i}^2, 
M_{\tilde b_j}^2)\nonumber\\
&& \simeq -\frac{1}{2 M_S^2} f_1(R_t, R_b) - \frac{m_t^2}{24
M_S^4}f_2 (R_t,R_b) - \frac{M_{H^+}^2}{24 M_S^4}f_3 (R_t,R_b) - \frac{m_b^2}{24
M_S^4}f_4 (R_t,R_b)\nn\\
&& \quad-(-1)^i \frac{\Delta_t^2}{6 M_S^4} f_5 (R_t,R_b)-
(-1)^j \frac{\Delta_b^2}{6 M_S^4} f_6 (R_t,R_b)\nn\\
&&\quad - \frac{\Delta_t^4 }{12 M_S^6}f_7 (R_t,R_b)
 - \frac{\Delta_b^4 }{12 M_S^6}f_8 (R_t,R_b) 
-(-1)^{i+j} \frac{\Delta_t^2 \Delta_b^2 }{12 M_S^6}f_9 (R_t,R_b)\,, \nn\\
\nn\\
&& C_{11}(m_t^2,M_{H^+}^2,m_b^2;M_{\tilde g}^2, M_{\tilde t_i}^2,
    M_{\tilde b_j}^2)\nonumber\\
&& \simeq \frac{1}{3 M_S^2} f_{10} (R_t,R_b)
 + (-1)^i \frac{\Delta_t^2}{8 M_S^4} f_{11} (R_t,R_b) + (-1)^j
 \frac{\Delta_b^2}{8 M_S^4} f_{12} (R_t,R_b)\,,\nn\\
\nn\\
&& C_{12}(m_t^2,M_{H^+}^2,m_b^2;M_{\tilde g}^2,M_{\tilde t_i}^2,
M_{\tilde b_j}^2)\nonumber \\
&&\simeq \frac{1}{6 M_S^2} f_{13}(R_t,R_b) 
        + (-1)^i \frac{\Delta_t^2}{24 M_S^4} f_{14}(R_t,R_b)
        + (-1)^j \frac{\Delta_b^2}{12 M_S^4} f_{15}(R_t,R_b)\,,\nn \\
\nn\\
&&B_0(m_q^2;M_{\tilde q_i}^2,M_{\tilde g}^2)
    \simeq \Delta - ln \frac{M_{\tilde g}^2}{R_q^2 \mu_0^2} + g_1 (R_q) 
     + (-1)^i \frac{\Delta_q^2}{2 M_S^2} f_1(R_q)\nn \\ 
&&\quad +  \frac{m_q^2}{6 M_S^2} f_2(R_q)
+ \frac{\Delta_q^4}{6 M_S^4} f_{2}(R_q) + (-1)^i 
    \frac{m_q^2 \Delta_q^2}{12 M_S^4} (2f_3-f_4)(R_q)\,,\nn \\
&&B_1(m_q^2;M_{\tilde q_i}^2,M_{\tilde g}^2)
    \simeq - \frac{1}{2} (\Delta - ln \frac{M_{\tilde g}^2}{R_q^2 \mu_0^2}) 
+ g_2 (R_q) 
     - (-1)^i \frac{\Delta_q^2}{6 M_S^2} f_2(R_q)\nn \\ 
&&\quad-  \frac{m_q^2}{12
    M_S^2} f_3(R_q) - \frac{\Delta_q^4}{24 M_S^4} (2f_3-f_4)(R_q)\,.
\eea

Where $M_S^2$ and $\Delta^2$
are defined in eq.~(\ref{eq:MSandDeltat}) and the functions $f_i$ and
$g_i$ are given in terms of the ratios
$R_t \equiv M_{\tilde g}/M_S$ and $R_b \equiv M_{\tilde g}/\tilde M_S$
as follows:\\
\bea
\label{eq:fyg}
f_1(R_1,R_2)&=&- \frac{2 R_2^2}{(1-R_2^2)(1-R_1^2)(R_2^2-R_1^2)}
\left[ ln \frac{R_1^2}{R_2^2} + R_1^2 lnR_2^2 - R_2^2 lnR_1^2 \right]
\nn \\
f_2(R_1,R_2)&=&\frac{12 R_2^2}{(1-R_2^2)^2(1-R_1^2)^3(R_2^2-R_1^2)^2}
\left[ R_1^2 - R_1^6 - R_2^2 -3 R_1^2 R_2^2 \right. \nn \\
&+&  R_1^6 R_2^2 + 3 R_1^4 R_2^2
+3 R_2^4 -3 R_1^4 R_2^4 -2 R_2^6 +2 R_1^2 R_2^6 + 2 R_1^4 lnR_1^2 \nn
\\
&+& R_2^2 ln\frac{R_2^2}{R_1^2} - R_1^2 R_2^2 lnR_1^2 
- 3 R_1^2 R_2^2
lnR_2^2 -R_1^6 R_2^2 lnR_2^2 - 4 R_1^4 R_2^2 lnR_1^2 \nn \\
&+& 2R_2^4 lnR_1^2
+2 R_1^2 R_2^4 lnR_1^2 + 2R_1^4 R_2^4 lnR_1^2 + 3R_1^4 R_2^2 lnR_2^2 \nn \\
&-& \left.  R_2^6 lnR_1^2 - 
R_1^2 R_2^6 lnR_1^2\right] \nn \\
f_3(R_1,R_2)&=& f_4(R_1,R_2)=\frac{12 R_2^4}{(1-R_2^2)^2(1-R_1^2)^2(R_1^2-R_2^2)^3}
\left[ 3R_1^4 - 2R_1^2 \right. \nn \\ 
&-& R_1^6 + 2 R_2^2
+  R_1^6 R_2^2 -  3 R_1^4 R_2^2 - 3 R_2^4  +  3 R_2^4 R_1^2 + R_2^6
- R_2^6 R_1^2 \nn \\ 
&+& 2 R_1^4 ln\frac{R_2^2}{R_1^2} 
- R_1^2
ln\frac{R_2^2}{R_1^2} -  R_1^6 lnR_2^2 - R_2^2 ln\frac{R_2^2}{R_1^2} + 2 R_2^2 R_1^2 
ln\frac{R_2^2}{R_1^2} \nn \\
&-& R_2^2 R_1^4 lnR_2^2+ 4 R_2^2 R_1^4 lnR_1^2 
 + 2 R_2^4 ln\frac{R_2^2}{R_1^2} - 4 R_2^4 R_1^2 lnR_2^2 \nn \\
&+& R_2^4 R_1^2 lnR_1^2
+ \left.  2 R_2^4 R_1^4 
ln\frac{R_2^2}{R_1^2}  + R_2^6 lnR_1^2 \right] \nn \\
f_5(R_1,R_2)&=& - \frac{6 R_2^2}{(1-R_2^2) (1-R_1^2)^2(R_1^2-R_2^2)^2}
\left[ - R_1^2 + R_1^4 + R_2^2  -  R_1^4 R_2^2 \right. \nn \\
&-& R_2^4 + R_1^2 R_2^4
- R_1^4 lnR_1^2 - R_2^2 ln\frac{R_2^2}{R_1^2} + 2 R_2^2 R_1^2
lnR_2^2 \nn \\
&-& \left. R_2^2 R_1^4 ln\frac{R_2^2}{R_1^2} -  R_2^4 lnR_1^2 \right] \nn \\
f_6(R_1,R_2)&=& \frac{6 R_2^4}{R_1^2 (1-R_2^2)^2 (1-R_1^2)(R_1^2-R_2^2)^2}
\left[ - R_1^2 + R_1^4 + R_2^2  -  R_1^4 R_2^2 \right. \nn \\
&-& R_2^4 + R_1^2 R_2^4 + R_1^4 lnR_2^2 - R_1^2 ln\frac{R_2^2}{R_1^2} - 2 R_2^2 R_1^2
lnR_1^2 \nn \\
&-&\left.  R_2^4 R_1^2 ln\frac{R_2^2}{R_1^2} +  R_2^4 lnR_2^2 \right] \nn \\
f_7(R_1,R_2)&=& \frac{6 R_2^2}{(1-R_2^2) (1-R_1^2)^3(R_2^2-R_1^2)^3}
\left[-R_1^4 + R_1^8 +4 R_2^2 R_1^2 -3 R_2^2 R_1^4 \right. \nn \\
&-&  R_2^2 R_1^8
- 3 R_2^4 +3 R_2^4 R_1^4 + 3 R_2^6 + R_2^6 R_1^4 -4 R_2^6 R_1^2
-2 R_1^6 lnR_1^2 \nn \\
&+& 6 R_2^2 R_1^4 lnR_1^2 -  6 R_2^4 R_1^2 lnR_2^2 
+ 2R_2^4 ln\frac{R_2^2}{R_1^2} +6 R_2^4 R_1^4 ln\frac{R_2^2}{R_1^2} 
- 2 R_2^4 R_1^6 ln\frac{R_2^2}{R_1^2} \nn \\
&+& \left.   2 R_2^6 lnR_1^2 \right] \nn \\
f_8(R_1,R_2)&=& \frac{6 R_2^6}{R_1^4(1-R_2^2)^3 (1-R_1^2)(R_2^2-R_1^2)^3}
\left[3R_1^4 -3R_1^6 -4 R_2^2 R_1^2 +4 R_2^2 R_1^6 \right. \nn \\
&+& 3 R_2^4R_1^2 
+ R_2^4  
- 3 R_2^4 R_1^4 -R_2^4 R_1^6 -R_2^8 +  R_2^8 R_1^2 + 2  R_1^4
ln\frac{R_2^2}{R_1^2} \nn \\
&-&2 R_1^6 lnR_2^2 +6 R_2^2 R_1^4 lnR_1^2 
- 6 R_1^2 R_2^4 lnR_2^2 + 6 R_2^4 R_1^4 ln\frac{R_2^2}{R_1^2} \nn \\
&+& \left.  2 R_2^6 lnR_2^2 -2
R_2^6 R_1^4 ln\frac{R_2^2}{R_1^2} \right]\nn \\
f_9(R_1,R_2)&=& \frac{12 R_2^4}{R_1^2(1-R_2^2)^2 (1-R_1^2)^2(R_1^2-R_2^2)^3}
\left[  R_2^6 +  R_1^4 -  R_1^6 +3 R_2^2 R_1^6 \right. \nn \\
&-& 2 R_2^4 R_1^6 +3 R_2^4
R_1^2 - 3 R_2^2 R_1^4 +2 R_2^6 R_1^4 -3 R_2^6 R_1^2 -  R_2^4 
-3 R_2^2 R_1^4 lnR_1^2 \nn \\
&+& R_1^6 lnR_1^2 - R_2^6 lnR_2^2
-  R_2^4 R_1^2 ln\frac{R_2^2}{R_1^2} + 3 R_2^4 R_1^2 lnR_1^2 
+2  R_2^2 R_1^2 ln\frac{R_2^2}{R_1^2} \nn \\
&+&2  R_2^6 R_1^2
ln\frac{R_2^2}{R_1^2} + 2  R_2^4 R_1^4 ln\frac{R_2^2}{R_1^2} +  
2  R_2^2 R_1^6 ln\frac{R_2^2}{R_1^2} - R_2^4 R_1^6
ln\frac{R_2^2}{R_1^2} \nn \\
&-& \left. 4 R_2^2 R_1^4 ln\frac{R_2^2}{R_1^2} - R_2^6 R_1^4
ln\frac{R_2^2}{R_1^2} \right] \nn \\
f_{10}(R_1,R_2)&=& \frac{3 R_2^2}{2(1-R_2^2)^2 (1-R_1^2)^2(R_1^2-R_2^2)}
\left[-R_1^2+R_1^4+R_2^2-R_2^2R_1^4 \right. \nn \\
&-& R_2^4 + R_1^2R_2^4 
+  2 R_1^2 ln\frac{R_2^2}{R_1^2} - ln\frac{R_2^2}{R_1^2} - R_1^4
lnR_2^2 + 2 R_2^2 ln\frac{R_2^2}{R_1^2} \nn \\
&+& \left.  2 R_2^2R_1^4 lnR_2^2 - 
4 R_2^2 R_1^2 ln\frac{R_2^2}{R_1^2} 
+  R_2^4 lnR_1^2 - 2 R_2^4R_1^2
lnR_1^2 \right] \nn \\
f_{11}(R_1,R_2)&=& - \frac{4 R_2^2}{(1-R_2^2)^2(1-R_1^2)^3(R_2^2-R_1^2)^2}
\left[- R_1^2 +4R_1^4-3R_1^6 + R_2^2 \right. \nn \\
&-& 3R_1^2R_2^2-3R_1^4R_2^2  
+ 5 R_1^6R_2^2 - R_2^4 + 6R_1^2R_2^4 -3R_1^4R_2^4 -2 R_1^6R_2^4 \nn \\
&-& 2 R_2^6R_1^2
+ 2 R_2^6R_1^4 + 2 R_1^6 lnR_1^2 - R_2^2 ln\frac{R_2^2}{R_1^2} 
+3 R_2^2 R_1^2 ln\frac{R_2^2}{R_1^2} \nn \\ 
&-&3 R_2^2 R_1^4lnR_2^2 
+ R_2^2 R_1^6lnR_2^2 -4 R_2^2 R_1^6lnR_1^2 +2 R_2^4 ln\frac{R_2^2}{R_1^2} 
-6 R_2^4 R_1^2 ln\frac{R_2^2}{R_1^2}\nn \\
&+& \left. 6 R_2^4 R_1^4lnR_2^2 - 2 R_2^4 R_1^6 ln\frac{R_2^2}{R_1^2} + R_2^6
lnR_1^2 -3R_2^6 R_1^2lnR_1^2 \right] \nn \\
f_{12}(R_1,R_2)&=& \frac{4 R_2^4}{R_1^2(1-R_2^2)^3(1-R_1^2)^2(R_1^2-R_2^2)^2}
\left[3R_1^2R_2^4 + 2R_1^6R_2^2 + 3R_1^2R_2^2 \right. \nn \\
&-& 6R_1^4R_2^2 +
3R_1^4R_2^4 - 2R_1^6R_2^4 - 5R_1^2R_2^6 + 2R_1^4R_2^6 - R_1^2 + R_2^2
+ R_1^4 \nn \\
&-& 4R_2^4 + 3R_2^6 + 3R_2^4 R_1^2lnR_1^2 
-4R_2^4 R_1^2lnR_2^2 -
6R_2^2 R_1^4lnR_2^2 \nn \\
&+& 6R_2^2 R_1^4lnR_1^2 + 
3R_2^2 R_1^6lnR_2^2 + 4R_2^4 R_1^2lnR_2^2 
- 6R_2^4 R_1^4lnR_1^2 \nn \\
&+& 4R_2^6 R_1^2lnR_2^2 - R_2^6 R_1^2lnR_1^2 - 2R_2^6lnR_2^2 - 
2R_2^6 R_1^4 ln\frac{R_2^2}{R_1^2} \nn \\
&+& \left.  3R_2^2 R_1^2
ln\frac{R_2^2}{R_1^2} 
+ 2R_1^4 ln\frac{R_2^2}{R_1^2} - 
R_1^2 ln\frac{R_2^2}{R_1^2} - R_1^6lnR_2^2\right]\nn \\
f_{13}(R_1,R_2)&=& -\frac{3 R_2^2}{(1-R_2^2)^2(1-R_1^2)(R_1^2-R_2^2)^2}
\left[R_1^4 - R_1^2 + R_2^2 - R_1^4R_2^2 - R_2^4 \right. \nn \\
&+& R_1^2R_2^4
+  R_2^2 R_1^4lnR_2^2 - 2R_2^4 R_1^2lnR_2^2 + R_2^6lnR_1^2 -
R_2^2ln\frac{R_2^2}{R_1^2} \nn \\
&+& \left. 2R_2^4 ln\frac{R_2^2}{R_1^2} \right] \nn \\
f_{14}(R_1,R_2)&=& \frac{12 R_2^4}{(1-R_2^2)^2(1-R_1^2)^2(R_1^2-R_2^2)^3}
\left[3R_1^4 - 2R_1^2 - R_1^6 + 2R_2^2 
+ R_1^6R_2^2 \right. \nn \\
&-& 3R_1^4R_2^2 + 3R_2^4 
+ 3R_1^2R_2^4 + R_2^6 - R_1^2R_2^6 - R_1^6lnR_2^2 + 4R_2^2
R_1^4lnR_1^2 \nn \\
&-& R_2^2 R_1^4lnR_2^2 + R_2^4 R_1^2lnR_1^2 
- 4R_2^4
R_1^2lnR_2^2 
+ R_2^6lnR_1^2 + 2R_1^4 ln\frac{R_2^2}{R_1^2} \nn \\
&-& \left. R_1^2
ln\frac{R_2^2}{R_1^2} -  R_2^2 ln\frac{R_2^2}{R_1^2}
+ 2R_2^2 R_1^2 ln\frac{R_2^2}{R_1^2} 
+ 2R_2^4 R_1^4
ln\frac{R_2^2}{R_1^2} + 2R_2^4ln\frac{R_2^2}{R_1^2} \right] \nn \\
f_{15}(R_1,R_2)&=& \frac{6R_2^4 }{R_1^2(1-R_2^2)^3(1-R_1^2)(R_1^2-R_2^2)^3}
\left[ -R_1^6 + R_1^4 - R_2^4 - 3R_1^2R_2^4 \right. \nn \\
&+& R_1^6R_2^4 + 3R_1^4R_2^4
+ 4R_2^6 - 4R_1^4R_2^6 - 3R_2^8 + 3R_1^2R_2^8 - 2R_2^2 R_1^6lnR_2^2 \nn \\
&+& 6R_2^4 R_1^4lnR_2^2 - 6R_2^6 R_1^2lnR_1^2 +2R_2^8lnR_2^2 
+ 2R_2^2 R_1^2 ln\frac{R_2^2}{R_1^2} \nn \\
&-& \left.  6R_2^4 R_1^2 ln\frac{R_2^2}{R_1^2}
- 2R_2^8 R_1^2 ln\frac{R_2^2}{R_1^2} \right] \nn \\
f_1 (R) &=& f_1 (R,R) = \frac{2}{(1-R^2)^2}\left[1 - R^2 + R^2 lnR^2 \right] \nn\\
f_{2}(R) &=& \frac{3}{(1-R^2)^3}\left[1 - R^4 + 2R^2 lnR^2 \right] \nn\\
f_{3}(R) &=& \frac{2}{(1-R^2)^4}\left[2 + R^6 - 6R^4 + 3R^2 + 6R^2 lnR^2 \right] \nn\\
f_{4}(R) &=& \frac{2}{(1-R^2)^4}\left[1- 6R^2 + 3R^4 + 2R^6   
- 6R^4 lnR^2\right] \nn\\
g_1(R) &=& \frac{1}{(1-R^2)} \left[1 - R^2 + R^2 lnR^2 \right] \nn \\
g_2(R) &=& \frac{1}{4(1-R^2)^2} \left[-3 + 4R^2 - R^4 - 4R^2 lnR^2 +
2R^4 lnR^2 \right]\,. \nn \\
\eea
Note that in the special case $M_{\tilde g} = M_S = \tilde M_S$,
$R_t=R_b=1$, the functions above are normalized as $f_i(1,1)=1$,
$f_i(1)=1$ and $g_i(1)=0$. Notice also that we use here the same notation
as in ref~\cite{HaberTemes}. 

\subsection*{{\bf{B.2}} Near-zero sbottom and stop  mixing}\label{ap:min}
The loop integrals are expanded as follows:
\bea
\label{eq:minbt}
 &&C_0(m_t^2,M_{H^+}^2,m_b^2;M_{\tilde g}^2, M_{\tilde t_i}^2,
    M_{\tilde b_j}^2) 
\simeq -\frac{1}{2M_{\tilde t_i}^2} f_1(R_{t_i},R_{b_j}) \nn \\
 &&\quad - \frac{m_{t}^2}{24 M_{\tilde t_i}^4} f_2(R_{t_i},R_{b_j}) 
- \frac{M_{H^+}^2}{24 M_{\tilde t_i}^4} f_3(R_{t_i},R_{b_j})
        - \frac{m_{b}^2}{24 M_{\tilde t_i}^4} f_4(R_{t_i},R_{b_j})\,,\nn \\
    &&C_{11}(m_t^2,M_{H^+}^2,m_b^2;M_{\tilde g}^2, M_{\tilde t_i}^2,
    M_{\tilde b_j}^2)
    \simeq \frac{1}{3M_{\tilde t_i}^2} f_{10}(R_{t_i}, R_{b_j})\,, \nn \\
    &&C_{12}(m_t^2,M_{H^+}^2,m_b^2;M_{\tilde g}^2, M_{\tilde t_i}^2,
    M_{\tilde b_j}^2)
    \simeq \frac{1}{6M_{\tilde t_i}^2} f_{13}(R_{t_i},R_{b_j})\,, \nn \\
   &&B_0(m_q^2;M_{\tilde q_i}^2,M_{\tilde g}^2)
    \simeq \Delta - ln \frac{M_{\tilde q_i}^2}{\mu_0^2} + g_1 (R_{q_i}) 
     +  \frac{m_q^2}{6 M_{\tilde q_i}^2} f_2(R_{q_i})\,, \nn \\
    &&B_1(m_q^2;M_{\tilde q_i}^2,M_{\tilde g}^2)
    \simeq - \frac{1}{2} (\Delta - ln \frac{M_{\tilde q_i}^2}{\mu_0^2}) 
+ g_2 (R_{q_i}) -  \frac{m_q^2}{12 M_{\tilde q_i}^2} f_3(R_{q_i})\,.\nn\\
\eea
Where $R_{q_i} \equiv M_{\tilde g}/M_{\tilde q_i}$ ($i=1,2$) and the
functions $f_i$ and $g_i$ are the same as in Section B.1.

\subsection*{{\bf{B.3}} Maximal sbottom mixing and near-zero stop mixing}

The loop integrals are expanded as follows:
\bea
    && C_0(m_t^2,M_{H^+}^2,m_b^2;M_{\tilde g}^2, M_{\tilde t_i}^2,
    M_{\tilde b_j}^2)
\simeq -\frac{1}{2 M_{\tilde t_i}^2} f_1(R_{t_i}, R_b)\nonumber \\
    &&\quad -\frac{m_t^2}{24 M_{\tilde t_i}^4}f_2 (R_{t_i},R_b) -
    \frac{M_{H^+}^2}{24 M_{\tilde t_i}^4} f_3 (R_{t_i},R_b)
- \frac{m_b^2}{24 M_{\tilde t_i}^4} f_4 (R_{t_i},R_b)\nn\\
&&\quad- (-1)^j \frac{\Delta_b^2}{6 M_{\tilde t_i}^4} f_6 (R_{t_i},R_b)- 
\frac{\Delta_b^4 }{12 M_{\tilde t_i}^6}f_8 (R_{t_i},R_b)\,, \nn\\
&& C_{11}(m_t^2,M_{H^+}^2,m_b^2;M_{\tilde g}^2, M_{\tilde t_i}^2,
    M_{\tilde b_j}^2)
\simeq \frac{1}{3 M_{\tilde t_i}^2} f_{10} (R_{t_i},R_b)
 + (-1)^j \frac{\Delta_b^2}{8 M_{\tilde t_i}^4} f_{12} (R_{t_i},R_b),\nn \\
&& C_{12}(m_t^2,M_{H^+}^2,m_b^2;M_{\tilde g}^2,M_{\tilde t_i}^2,
    M_{\tilde b_j}^2)
   \simeq \frac{1}{6 M_{\tilde t_i}^2} f_{13}(R_{t_i},R_b) 
        + (-1)^j \frac{\Delta_b^2}{12 M_{\tilde t_i}^4} f_{15}(R_{t_i},R_b).
\nn\\
\eea
Here $R_{t_i} \equiv M_{\tilde g}/M_{\tilde t_i}$ ($i=1,2$);  $R_b
    \equiv M_{\tilde g}/\tilde M_S$ and the functions $f_i$ are the
same as in Section B.1.
For the two-point integrals $B_{0,1}(m_b^2,M_{\tilde g}^2,M_{\tilde b_i}^2)$ 
and $B_{0,1}(m_t^2,M_{\tilde g}^2,M_{\tilde t_i}^2)$ 
we use the expressions given in~(\ref{eq:maxbt})
and~(\ref{eq:minbt}) respectively.

\subsection*{{\bf{B.4}} Maximal stop mixing and near-zero sbottom mixing}

The loop integrals are expanded as follows:
\bea
&& C_0(m_t^2,M_{H^+}^2,m_b^2;M_{\tilde g}^2, M_{\tilde t_i}^2,
    M_{\tilde b_j}^2)\nonumber \\
&& \simeq -\frac{1}{2 M_S^2} f_1(R_t, R_{b_j}) - \frac{m_t^2}{24
M_S^4}f_2 (R_t,R_{b_j}) - \frac{M_{H^+}^2}{24 M_S^4}f_3 (R_t,R_{b_j})
- \frac{m_b^2}{24 M_S^4}f_4 (R_t,R_{b_j})\nn \\
        &&\quad - (-1)^i \frac{\Delta_t^2}{6 M_S^4} f_5 (R_t,R_{b_j})
- \frac{\Delta_t^4 }{12 M_S^6}f_7 (R_t,R_{b_j})\,, \nn \\
&& C_{11}(m_t^2,M_{H^+}^2,m_b^2;M_{\tilde g}^2, M_{\tilde t_i}^2,
    M_{\tilde b_j}^2)
\simeq \frac{1}{3 M_S^2} f_{10} (R_t,R_{b_j})
 + (-1)^i \frac{\Delta_t^2}{8 M_S^4} f_{11} (R_t,R_{b_j}), \nn \\
&& C_{12}(m_t^2,M_{H^+}^2,m_b^2;M_{\tilde g}^2,M_{\tilde t_i}^2,
    M_{\tilde b_j}^2)
\simeq \frac{1}{6 M_S^2} f_{13}(R_t,R_{b_j}) 
        + (-1)^i \frac{\Delta_t^2}{24 M_S^4} f_{14}(R_t,R_{b_j}), \nn \\
\eea
where $R_{b_i} \equiv M_{\tilde g}/M_{\tilde b_i}$ ($i=1,2$); 
$R_t \equiv M_{\tilde g}/M_S$ and again the functions $f_i$ are
        as in Section B.1 . In this case for the two-point integrals 
$B_{0,1}(m_b^2,M_{\tilde g}^2,M_{\tilde
    b_i}^2)$ and $B_{0,1}(m_t^2,M_{\tilde g}^2,M_{\tilde t_i}^2)$
we use the expressions given in~(\ref{eq:minbt}) and~(\ref{eq:maxbt})
respectively.

\section*{{\bf{C}} Complete expressions for $\Delta_{SQCD}$ to order 
$M_{EW}^2/M_{SUSY}^2$}
\renewcommand{\theequation}{C.\arabic{equation}}

Here we give the complete expressions for the expansion of the SUSY-QCD
correction to $\Gamma (H^+ \rightarrow t \bar b)$ up to order
$M_{EW}^2/M_{SUSY}^2$. From eqs.~(\ref{eq:CTdecay}),~(\ref{eq:defSQCD}) and~(\ref{eq:loopsdecay}) 
the SUSY-QCD correction can be rewritten as:
\bea
\Delta_{SQCD} = \frac{U_t}{D} (K_t + H_t) +  \frac{U_b}{D} (K_b + H_b)\,,\nn\\
\eea
where we have defined:
\bea
K_t = \frac{\delta m_t}{m_t}
        +\frac{1}{2}\,\delta Z_L^b+\frac{1}{2}\,\delta Z_R^t\,, \nn\\
K_b = \frac{\delta m_b}{m_b}
        +\frac{1}{2}\,\delta Z_L^t+\frac{1}{2}\,\delta Z_R^b\,.
\eea

In the following we give the expressions for $K_t$, $H_t$, $K_b$ and
$H_b$ up to order
$M_{EW}^2/M_{SUSY}^2$ in the four different cases that we have
considered in this paper.

\subsection*{{\bf{C.1}} Maximal stop and sbottom mixing}

The results for $K_t$, $H_t$, $K_b$ and
$H_b$ up to order $M_{EW}^2/M_{SUSY}^2$ are the following:
\bea
K_t &=& \frac{\alpha_s}{3 \pi}\,\left[\,M_{\tilde g}\,
\left( \frac{X_t}{M_S^2}\,f_1(R_t)-
\frac{m_b^2 X_b}{6\,\tilde M_S^4}\,(2f_3-f_4)(R_b)\right) 
-\frac{1}{2}\, ln \frac{R_b^2}{R_t^2} + g_1(R_t) -g_1(R_b) \right. \nn \\
&+& g_2(R_t) - g_2(R_b)
-\frac{m_t^2}{12 M_S^2}\,(2 f_2-f_3)(R_t) - \frac{m_b^2}{4 \tilde M_S^2}\,(2f_2-f_3)(R_b)\,\nn\\
&+&  \frac{m_t^2 X_t^2}{24
M_S^4}(4f_2-2f_3+f_4)(R_t) -\frac{m_b^2 X_b^2}{24 \tilde M_S^4}
(4f_2-2f_3+f_4)(R_b)\nn\\
&-& \left. \frac{M_L^2-M_R^2}{12 M_S^2}(3f_1-f_2)(R_t)
+ \frac{M_L^{'2}-M_R^{'2}}{12 \tilde M_S^2}(3f_1-f_2)(R_b)\right]\,,  \nn\\
H_t &=& - \frac{2\alpha_s}{3\pi} \frac{1}{m_t \cot\beta}\left[
-\frac{M_{\tilde g} g_{RL}}{2 M_S^2}f_1 (R_t,R_b) +\frac{m_t g_{LL}}{6
M_S^2} (2 f_{10} - f_{13})(R_t,R_b)\right. \nn \\ 
&+& \frac{M_{\tilde g} g_{LL}}{6 M_S^4}m_t X_t f_5 (R_t,R_b)
+\left. \frac{m_b g_{RR}}{6 M_S^2} f_{13}(R_t,R_b)  + \frac{M_{\tilde
g} g_{RR}}{6 M_S^4}m_b X_b f_6 (R_t,R_b)\right. \nn \\
&-& \frac{m_t m_b X_b g_{LR}}{24
M_S^4} (3 f_{12} -2 f_{15})(R_t,R_b)
-\frac{m_t m_b X_t g_{LR}}{24
M_S^4} f_{14}(R_t,R_b)\nn \\
&-&\left. \frac{M_{\tilde g} m_t m_b X_t X_b g_{LR}}{12
M_S^6}f_9(R_t,R_b) + \frac{m_t^2 X_t
g_{RL}}{24 M_S^4}(f_{14}-3f_{11})(R_t,R_b) \right. \nn \\
&-&\left.\frac{m_b^2 X_b
g_{RL}}{12 M_S^4}f_{15}(R_t,R_b)-\frac{M_{\tilde g} g_{RL}}{24
M_S^4}(m_t^2 f_2 + M_{H^+}^2 f_3 + m_b^2 f_4)(R_t,R_b)\right. \nn \\
&-&\frac{M_{\tilde g} m_t^2 X_t^2 g_{RL}}{12 M_S^6}f_7 (R_t,R_b)-
\frac{M_{\tilde g} m_b^2 X_b^2 g_{RL}}{12 M_S^6}f_8 (R_t,R_b)\nn\\
&+&\left.
\frac{M_{\tilde g} (M_L^{'2}-M_R^{'2}) g_{RL}}{12 M_S^4}f_6 (R_t,R_b)-
\frac{M_{\tilde g} (M_L^{2}-M_R^{2}) g_{RL}}{12 M_S^4}f_5 (R_t,R_b)
\right]\,,\nn\\ 
K_b &=& \frac{\alpha_s}{3 \pi}\,\left[\,M_{\tilde g}\,
\left( \frac{X_b}{\tilde M_S^2}\,f_1(R_b)-
\frac{m_t^2 X_t}{6\,M_S^4}\,(2f_3-f_4)(R_t)\right) 
+\frac{1}{2}\, ln \frac{R_b^2}{R_t^2} - g_1(R_t) +g_1(R_b) \right. \nn \\
&-& g_2(R_t) + g_2(R_b)
-\frac{m_b^2}{12 \tilde M_S^2}\,(2 f_2-f_3)(R_b) - \frac{m_t^2}{4  M_S^2}\,(2f_2-f_3)(R_t)\,\nn\\
&-&  \frac{m_t^2 X_t^2}{24
M_S^4}(4f_2-2f_3+f_4)(R_t) +\frac{m_b^2 X_b^2}{24 \tilde M_S^4}
(4f_2-2f_3+f_4)(R_b)\nn\\
&+& \left. \frac{M_L^2-M_R^2}{12 M_S^2}(3f_1-f_2)(R_t)
- \frac{M_L^{'2}-M_R^{'2}}{12 \tilde M_S^2}(3f_1-f_2)(R_b)\right]\,,  \nn\\
H_b &=& - \frac{2\alpha_s}{3\pi} \frac{1}{m_b \tan\beta}\left[
-\frac{M_{\tilde g} g_{LR}}{2 M_S^2}f_1 (R_t,R_b) +\frac{m_t g_{RR}}{6
M_S^2} (2 f_{10} - f_{13})(R_t,R_b)\right. \nn \\ 
&+& \frac{M_{\tilde g} g_{RR}}{6 M_S^4}m_t X_t f_5 (R_t,R_b)
+\frac{m_b g_{LL}}{6 M_S^2} f_{13}(R_t,R_b) 
+\frac{M_{\tilde g} g_{LL}}{6 M_S^4}m_b X_b f_6 (R_t,R_b)\nn\\
&-&\frac{m_t m_b X_b g_{RL}}{24
M_S^4} (3 f_{12} -2 f_{15})(R_t,R_b)-\frac{m_t m_b X_t g_{RL}}{24
M_S^4} f_{14}(R_t,R_b)\nn \\
&-&\left. \frac{M_{\tilde g} m_t m_b X_t X_b g_{RL}}{12
M_S^6}f_9(R_t,R_b) + \frac{m_t^2 X_t
g_{LR}}{24 M_S^4}(f_{14}-3f_{11})(R_t,R_b) \right. \nn \\
&-&\left. \frac{m_b^2 X_b
g_{LR}}{12 M_S^4}f_{15}(R_t,R_b)-\frac{M_{\tilde g} g_{LR}}{24
M_S^4}(m_t^2 f_2 + M_{H^+}^2 f_3 + m_b^2 f_4)(R_t,R_b)\right. \nn \\
&-&\frac{M_{\tilde g} m_t^2 X_t^2 g_{LR}}{12 M_S^6}f_7 (R_t,R_b)
-\frac{M_{\tilde g} m_b^2 X_b^2 g_{LR}}{12 M_S^6}f_8 (R_t,R_b)\nn\\
&-&\left.
\frac{M_{\tilde g} (M_L^{'2}-M_R^{'2}) g_{LR}}{12 M_S^4}f_6 (R_t,R_b)+
\frac{M_{\tilde g} (M_L^{2}-M_R^{2}) g_{LR}}{12 M_S^4}f_5 (R_t,R_b)
\right]\,, 
\eea

\subsection*{{\bf{C.2}} Near-zero stop and sbottom mixing}

The results for $K_t$, $H_t$, $K_b$ and
$H_b$ up to order $M_{EW}^2/M_{SUSY}^2$ are:
\begin{eqnarray}
K_t &=& \frac{\alpha_s}{3 \pi} \, \left[- \frac{2 M_{\tilde g}
X_t}{M_{\tilde t_1}^2-M_{\tilde t_2}^2} \left(ln\frac{M_{\tilde
t_2}^2}{M_{\tilde t_1}^2}+g_1 (R_{t_1}) - g_1 (R_{t_2})\right) 
-\frac{1}{2}\, ln \frac{M_{\tilde t_1}^2}{M_{\tilde b_1}^2}+
g_1(R_{t_1}) \right. \nn \\
&-& g_1(R_{b_1}) + g_2(R_{t_1}) 
- g_2(R_{b_1}) +\frac{m_t^2 X_t^2}{(M_{\tilde t_1}^2-M_{\tilde t_2}^2)^2}
\left(-\frac{1}{2}\, ln \frac{M_{\tilde t_2}^2}{M_{\tilde t_1}^2}+
g_1(R_{t_2}) -g_1(R_{t_1})\right. \nn \\
 &+& \left.  g_2(R_{t_2}) 
- g_2(R_{t_1}) \right) 
-\frac{m_b^2 X_b^2}{(M_{\tilde b_1}^2-M_{\tilde b_2}^2)^2}
\left( -\frac{1}{2}\, ln \frac{M_{\tilde b_2}^2}{M_{\tilde b_1}^2}+
g_1(R_{b_2}) -g_1(R_{b_1}) + g_2(R_{b_2}) \right. \nn \\
&-& \left.  g_2(R_{b_1}) \right) -\frac{m_b^2}{6 M_{\tilde b_1}^2}
(2f_2-f_3)(R_{b_1})
-\frac{m_t^2}{12 M_{\tilde t_2}^2} (2 f_2-f_3)(R_{t_2}) \nn \\
&-& \left. \frac{m_b^2}{12 M_{\tilde b_2}^2}
(2f_2-f_3)(R_{b_2})
+\frac{M_{\tilde g}m_b^2 X_b}{M_{\tilde b_1}^2-M_{\tilde
b_2}^2} \left( \frac{f_2(R_{b_1})}{3 M_{\tilde b_1}^2}
-\frac{f_2(R_{b_2})}
{3 M_{\tilde b_2}^2} \right) \right]\,,\nn\\
H_t &=& -  \frac{2\alpha_s}{3\pi} \frac{1}{m_t \cot \beta} \left[
-\frac{M_{\tilde g} g_{RL}}{2 M_{\tilde t_2}^2}f_1 (R_{t_2},R_{b_1}) 
+\frac{m_t g_{LL}}{6
M_{\tilde t_1}^2} (2 f_{10} - f_{13})(R_{t_1},R_{b_1})\right. \nn\\ 
&+&\frac{M_{\tilde g} m_t X_t}{M_{\tilde t_1}^2-M_{\tilde t_2}^2}
\left(g_{LL}+\frac{m_t X_t}{M_{\tilde t_1}^2-M_{\tilde t_2}^2}g_{RL}\right)
\left(\frac{f_1 (R_{t_2},R_{b_1})}{2M_{\tilde t_2}^2}- \frac{f_1
(R_{t_1},R_{b_1})}{2M_{\tilde t_1}^2} \right) \nn \\
&+& \frac{M_{\tilde g} m_b X_b}{M_{\tilde b_1}^2-M_{\tilde b_2}^2} 
\left(g_{RR}-\frac{m_b X_b}{M_{\tilde b_1}^2-M_{\tilde b_2}^2}g_{RL}\right)
\left(\frac{f_1 (R_{t_2},R_{b_2})}{2M_{\tilde t_2}^2}- \frac{f_1
(R_{t_2},R_{b_1})}{2M_{\tilde t_2}^2} \right) \nn \\
&-& \frac{m_t m_b X_b g_{LR}}{6 M_{\tilde t_1}^2
(M_{\tilde b_1}^2-M_{\tilde b_2}^2)} 
\left(f_{13}(R_{t_1},R_{b_1})-f_{13}(R_{t_1},R_{b_2})
-2 (f_{10}(R_{t_1},R_{b_1})-f_{10}(R_{t_1},R_{b_2}))\right)\nn \\
&-&\frac{m_t m_b X_t g_{LR}}{M_{\tilde t_1}^2-M_{\tilde t_2}^2} 
\left( \frac{f_{13}(R_{t_2},R_{b_2})}{6 M_{\tilde
t_2}^2}-\frac{f_{13}(R_{t_1},R_{b_2})}{6 M_{\tilde t_1}^2}\right)
+\frac{m_b g_{RR}}{6 M_{\tilde t_2}^2} f_{13}(R_{t_2},R_{b_2})\nn \\
&+&\frac{M_{\tilde g} m_t m_b X_t X_b g_{LR}}{(M_{\tilde
t_1}^2-M_{\tilde t_2}^2)(M_{\tilde b_1}^2-M_{\tilde b_2}^2)}\left(
\frac{f_1(R_{t_1},R_{b_2})}{2 M_{\tilde t_1}^2} +
\frac{f_1(R_{t_2},R_{b_1})}{2 M_{\tilde t_2}^2}-
\frac{f_1(R_{t_1},R_{b_1})}{2 M_{\tilde t_1}^2}-
\frac{f_1(R_{t_2},R_{b_2})}{2 M_{\tilde t_2}^2}\right)\nn \\ 
&-& \frac{m_t^2 X_t g_{RL}}{M_{\tilde t_1}^2-M_{\tilde t_2}^2}\left(
\frac{f_{13}(R_{t_1},R_{b_1})}{6 M_{\tilde t_1}^2} -
\frac{f_{10}(R_{t_1},R_{b_1})}{3 M_{\tilde t_1}^2} -
\frac{f_{13}(R_{t_2},R_{b_1})}{6 M_{\tilde t_2}^2}+
\frac{f_{10}(R_{t_2},R_{b_1})}{3 M_{\tilde t_2}^2} \right)\nn \\
&-&\left. \frac{m_b^2 X_b
g_{RL}}{6 M_{\tilde t_2}^2(M_{\tilde b_1}^2-M_{\tilde b_2}^2)}\left(f_{13}(R_{t_2},R_{b_2})-f_{13}(R_{t_2},R_{b_1})\right)\right. \nn \\
&-&\left.\frac{M_{\tilde g} g_{RL}}{24 M_{\tilde t_2}^4
}(m_t^2 f_2 + M_{H^+}^2 f_3 + m_b^2 f_4)(R_{t_2},R_{b_1})\right]\,,\nn \\
K_b &=&  \frac{\alpha_s}{3 \pi} \,\left[ - \frac{2 M_{\tilde g}
X_b}{M_{\tilde b_1}^2-M_{\tilde b_2}^2} \left( ln\frac{M_{\tilde
b_2}^2}{M_{\tilde b_1}^2}+g_1 (R_{b_1}) - g_1 (R_{b_2}) \right) 
-\frac{1}{2}\, ln \frac{M_{\tilde b_1}^2}{M_{\tilde t_1}^2}+
g_1(R_{b_1}) \right. \nn \\
&-& g_1(R_{t_1}) + g_2(R_{b_1}) 
- g_2(R_{t_1}) 
-\frac{m_t^2 X_t^2}{(M_{\tilde t_1}^2-M_{\tilde t_2}^2)^2}
\left(-\frac{1}{2}\, ln \frac{M_{\tilde t_2}^2}{M_{\tilde t_1}^2}+
g_1(R_{t_2}) -g_1(R_{t_1})\right. \nn \\
 &+& \left.  g_2(R_{t_2}) 
- g_2(R_{t_1}) \right) 
+\frac{m_b^2 X_b^2}{(M_{\tilde b_1}^2-M_{\tilde b_2}^2)^2}
\left( -\frac{1}{2}\, ln \frac{M_{\tilde b_2}^2}{M_{\tilde b_1}^2}+
g_1(R_{b_2}) -g_1(R_{b_1}) + g_2(R_{b_2}) \right. \nn \\
&-& \left.  g_2(R_{b_1}) \right) -\frac{m_t^2}{6 M_{\tilde t_1}^2}
(2f_2-f_3)(R_{t_1})
-\frac{m_b^2}{12 M_{\tilde b_2}^2} (2 f_2-f_3)(R_{b_2}) \nn \\
&-& \left. \frac{m_t^2}{12 M_{\tilde t_2}^2}
(2f_2-f_3)(R_{t_2})
+ \frac{M_{\tilde g}m_t^2 X_t}{M_{\tilde t_1}^2-M_{\tilde
t_2}^2} \left( \frac{f_2(R_{t_1})}{3 M_{\tilde t_1}^2}
-\frac{f_2(R_{t_2})}
{3 M_{\tilde t_2}^2} \right) \right]\,,\nn\\
H_b &=& - \frac{2\alpha_s}{3\pi} \frac{1}{m_b \tan\beta}\left[
-\frac{M_{\tilde g} g_{LR}}{2 M_{\tilde t_1}^2}f_1 (R_{t_1},R_{b_2}) 
+\frac{m_t g_{RR}}{6
M_{\tilde t_2}^2} (2 f_{10} - f_{13})(R_{t_2},R_{b_2})\right.\nn \\ 
&+&\frac{M_{\tilde g} m_t X_t}{M_{\tilde t_1}^2-M_{\tilde t_2}^2}
\left(g_{RR}-\frac{m_t X_t}{M_{\tilde t_1}^2-M_{\tilde t_2}^2}g_{LR}\right)
\left(\frac{f_1 (R_{t_2},R_{b_2})}{2 M_{\tilde t_2}^2}- \frac{f_1
(R_{t_1},R_{b_2})}{2 M_{\tilde t_1}^2}\right)\nn \\
&+& \frac{M_{\tilde g} m_b X_b}{M_{\tilde b_1}^2-M_{\tilde b_2}^2} 
\left(g_{LL}+\frac{m_b X_b}{M_{\tilde b_1}^2-M_{\tilde b_2}^2}g_{LR}\right)
\left(\frac{f_1 (R_{t_1},R_{b_2})}{2 M_{\tilde t_1}^2}- \frac{f_1
(R_{t_1},R_{b_1})}{2 M_{\tilde t_1}^2}\right)\nn \\
&-& \frac{m_t m_b X_b g_{RL}}{6 M_{\tilde t_2}^2
(M_{\tilde b_1}^2-M_{\tilde b_2}^2)} \left(f_{13}(R_{t_2},R_{b_1}) -2 f_{10}(R_{t_2},R_{b_1})-f_{13}(R_{t_2},R_{b_2})+2f_{10}(R_{t_2},R_{b_2})\right) \nn \\
&-&\frac{m_t m_b X_t g_{RL}}{M_{\tilde t_1}^2-M_{\tilde t_2}^2} 
\left( \frac{f_{13}(R_{t_2},R_{b_1})}{6 M_{\tilde
t_2}^2}-\frac{f_{13}(R_{t_1},R_{b_1})}{6 M_{\tilde t_1}^2}\right)
+\frac{m_b g_{LL}}{6 M_{\tilde t_1}^2} f_{13}(R_{t_1},R_{b_1})  \nn\\
&-&\frac{M_{\tilde g} m_t m_b X_t X_b g_{RL}}{(M_{\tilde
t_1}^2-M_{\tilde t_2}^2)(M_{\tilde b_1}^2-M_{\tilde b_2}^2)}\left(
\frac{f_1(R_{t_1},R_{b_1})}{2 M_{\tilde t_1}^2} -
\frac{f_1(R_{t_1},R_{b_2})}{2 M_{\tilde t_1}^2} -
\frac{f_1(R_{t_2},R_{b_2})}{2 M_{\tilde t_2}^2}+
\frac{f_1(R_{t_2},R_{b_1})}{2 M_{\tilde t_2}^2}\right)\nn\\ 
&-&\frac{m_t^2 X_t
g_{LR}}{M_{\tilde t_1}^2-M_{\tilde t_2}^2}\left(
\frac{f_{13}(R_{t_1},R_{b_2})}{6 M_{\tilde t_1}^2} -
\frac{f_{10}(R_{t_1},R_{b_2})}{3 M_{\tilde t_1}^2} -
\frac{f_{13}(R_{t_2},R_{b_2})}{6 M_{\tilde t_2}^2}+
\frac{f_{10}(R_{t_2},R_{b_2})}{3 M_{\tilde t_2}^2}\right)\nn\\
&-&\left.\frac{m_b^2 X_b
g_{LR}}{6 M_{\tilde t_1}^2(M_{\tilde b_1}^2-M_{\tilde b_2}^2)}
\left(f_{13}(R_{t_1},R_{b_2})-f_{13}(R_{t_1},R_{b_1})\right)\right. \nn \\
&-&\left.\frac{M_{\tilde g} g_{LR}}{24 M_{\tilde t_1}^4
}(m_t^2 f_2 + M_{H^+}^2 f_3 + m_b^2 f_4)(R_{t_1},R_{b_2})\right]\,,
\end{eqnarray}

\subsection*{{\bf{C.3}} Maximal sbottom mixing and near-zero stop mixing}

The results for $K_t$, $H_t$, $K_b$ and
$H_b$ up to order $M_{EW}^2/M_{SUSY}^2$ are:
\begin{eqnarray}
K_t &=& \frac{\alpha_s}{3 \pi}\,\left[ - \frac{2 M_{\tilde g}
X_t}{M_{\tilde t_1}^2-M_{\tilde t_2}^2} \left( ln\frac{M_{\tilde
t_2}^2}{M_{\tilde t_1}^2}+g_1 (R_{t_1}) - g_1 (R_{t_2}) \right) 
-\frac{1}{2}\, ln \frac{M_{\tilde t_1}^2}{\tilde M_S^2}+
g_1(R_{t_1}) \right. \nn \\
&-& g_1(R_{b}) + g_2(R_{t_1}) 
- g_2(R_{b}) +\frac{m_t^2 X_t^2}{(M_{\tilde t_1}^2-M_{\tilde t_2}^2)^2}
\left(-\frac{1}{2}\, ln \frac{M_{\tilde t_2}^2}{M_{\tilde t_1}^2}+
g_1(R_{t_2}) -g_1(R_{t_1})\right. \nn \\
 &+& \left.  g_2(R_{t_2}) 
- g_2(R_{t_1}) \right) 
-\frac{m_t^2}{12 M_{\tilde t_2}^2}(2f_2-f_3)(R_{t_2})
-\frac{m_b^2}{4 \tilde M_S^2}(2f_2-f_3)(R_{b}) \nn \\
&-& \left. \frac{m_b^2 X_b^2}{24 \tilde M_S^4}(4f_2 - 2f_3+f_4)(R_b)
+\frac{M_L^{'2}-M_R^{'2}}{12 \tilde M_S^2}(3f_1-f_2)(R_b) 
- \frac{M_{\tilde g}m_b^2 X_b}{6 \tilde M_S^4}
(2f_3-f_4)(R_b)\right]\,, \nn\\
H_t &=& - \frac{2\alpha_s}{3\pi} \frac{1}{m_t \cot\beta}\left[
-\frac{M_{\tilde g} g_{RL}}{2 M_{\tilde t_2}^2}f_1 (R_{t_2},R_b) 
+\frac{m_t g_{LL}}{6
M_{\tilde t_1}^2} (2 f_{10} - f_{13})(R_{t_1},R_b)\right. \nn \\ 
&+&\frac{M_{\tilde g} m_t X_t}{M_{\tilde t_1}^2-M_{\tilde t_2}^2}
\left(g_{LL}+\frac{m_t X_t}{M_{\tilde t_1}^2-M_{\tilde t_2}^2}g_{RL}\right)
\left(\frac{f_1 (R_{t_2},R_{b})}{2M_{\tilde t_2}^2}- \frac{f_1
(R_{t_1},R_{b})}{2M_{\tilde t_1}^2} \right) \nn \\
&+&\left. \frac{m_b g_{RR}}{6 M_{\tilde t_2}^2} f_{13}(R_{t_2},R_b)  +
\frac{M_{\tilde g} g_{RR}}{6 M_{\tilde t_2}^4} 
m_b X_b f_6 (R_{t_2},R_b)\right. \nn \\
&-& \frac{m_t m_b X_b g_{LR}}{24 M_{\tilde t_1}^4} \left(3
f_{12}(R_{t_1},R_b) -2 f_{15}(R_{t_1},R_b) \right)\nn \\
&-&\frac{m_t m_b X_t g_{LR}}{M_{\tilde t_1}^2-M_{\tilde t_2}^2} 
\left( \frac{f_{13}(R_{t_2},R_b)}{6 M_{\tilde
t_2}^2}-\frac{f_{13}(R_{t_1},R_b)}{6 M_{\tilde t_1}^2}\right)\nn\\
&+&\frac{M_{\tilde g} m_t m_b X_t X_b g_{LR}}{(M_{\tilde
t_1}^2-M_{\tilde t_2}^2)}\left(
\frac{f_6(R_{t_1},R_b)}{6 M_{\tilde t_1}^4} -
\frac{f_6(R_{t_2},R_b)}{6 M_{\tilde t_2}^4} \right)\nn\\
&-& \frac{m_t^2 X_t
g_{RL}}{M_{\tilde t_1}^2-M_{\tilde t_2}^2}\left(
\frac{f_{13}(R_{t_1},R_b)}{6 M_{\tilde t_1}^2} -
\frac{f_{10}(R_{t_1},R_b)}{3 M_{\tilde t_1}^2} -
\frac{f_{13}(R_{t_2},R_b)}{6 M_{\tilde t_2}^2}+
\frac{f_{10}(R_{t_2},R_b)}{3 M_{\tilde t_2}^2} \right)\nn \\
&-&\left.\frac{m_b^2 X_b
g_{RL}}{12 M_{\tilde t_2}^4}f_{15}(R_{t_2},R_b)
-\frac{M_{\tilde g} g_{RL}}{24 M_{\tilde t_2}^4
}(m_t^2 f_2 + M_{H^+}^2 f_3 + m_b^2 f_4)(R_{t_2},R_b)\right. \nn \\
&-&\left.\frac{M_{\tilde g} m_b^2 X_b^2 g_{RL}}{12 M_{\tilde
t_2}^6} f_8 (R_{t_2},R_b)+ \frac{M_{\tilde g}
(M_L^{'2}-M_R^{'2})g_{RL}}{12 M_{\tilde t_2}^4} f_6 (R_{t_2},R_b) 
\right]\,,\nn\\
K_b &=&  \frac{\alpha_s}{3 \pi}\,\left[ \frac{M_{\tilde g}X_b}
{\tilde M_S^2} f_1(R_b)
+\frac{1}{2}\, ln \frac{M_{\tilde t_1}^2}{\tilde M_S^2}-
g_1(R_{t_1}) + g_1(R_{b}) - g_2(R_{t_1}) 
+ g_2(R_{b}) \right. \nn \\
&-&\frac{m_t^2 X_t^2}{(M_{\tilde t_1}^2-M_{\tilde t_2}^2)^2}
\left(-\frac{1}{2}\, ln \frac{M_{\tilde t_2}^2}{M_{\tilde t_1}^2}+
g_1(R_{t_2}) -g_1(R_{t_1}) + g_2(R_{t_2}) - g_2(R_{t_1}) \right) \nn \\ 
&-&\frac{m_t^2}{12 M_{\tilde t_2}^2}(2f_2-f_3)(R_{t_2})
-\frac{m_b^2}{12 \tilde M_S^2}(2f_2-f_3)(R_{b}) \nn \\
&+&\frac{m_b^2 X_b^2}{24 \tilde M_S^4}(4f_2 - 2f_3+f_4)(R_b)
-\frac{M_L^{'2}-M_R^{'2}}{12 \tilde M_S^2}(3f_1-f_2)(R_b) \nn \\
&+& \left. \frac{M_{\tilde g}m_t^2 X_t}{M_{\tilde t_1}^2-M_{\tilde t_2}^2}
\left(\frac{f_2(R_{t_1})}{3 M_{\tilde t_1}^2} - 
\frac{f_2(R_{t_2})}{3 M_{\tilde t_2}^2} \right) 
- \frac{m_t^2}{6 M_{\tilde t_1}^2}(2f_2-f_3)(R_{t_1})\right]\,, \nn\\
H_b &=& - \frac{2\alpha_s}{3\pi} \frac{1}{m_b \tan\beta}\left[
-\frac{M_{\tilde g} g_{LR}}{2 M_{\tilde t_1}^2}f_1 (R_{t_1},R_b) 
+\frac{m_t g_{RR}}{6
M_{\tilde t_2}^2} (2 f_{10} - f_{13})(R_{t_2},R_b)\right.\nn \\ 
&+&\frac{M_{\tilde g} m_t X_t}{M_{\tilde t_1}^2-M_{\tilde t_2}^2}
\left(g_{RR}-\frac{m_t X_t}{M_{\tilde t_1}^2-M_{\tilde t_2}^2}g_{LR}\right)
\left(\frac{f_1 (R_{t_2},R_{b})}{2M_{\tilde t_2}^2}- \frac{f_1
(R_{t_1},R_{b})}{2M_{\tilde t_1}^2} \right) \nn \\
&+&\left. \frac{m_b g_{LL}}{6 M_{\tilde t_1}^2} f_{13}(R_{t_1},R_b)  
+ \frac{M_{\tilde
g} g_{LL}}{6 M_{\tilde t_1}^4} m_b X_b f_6 (R_{t_1},R_b)\right. \nn \\
&-& \frac{m_t m_b X_b g_{RL}}{24 M_{\tilde t_2}^4} 
\left(3 f_{12}(R_{t_2},R_b) -2 f_{15}(R_{t_2},R_b)\right)\nn \\
&-&\left.\frac{m_t m_b X_t g_{RL}}{M_{\tilde t_1}^2-M_{\tilde t_2}^2} 
\left( \frac{f_{13}(R_{t_2},R_b)}{6 M_{\tilde
t_2}^2}-\frac{f_{13}(R_{t_1},R_b)}{6 M_{\tilde t_1}^2}\right)\right.\nn \\ 
&+&\frac{M_{\tilde g} m_t m_b X_t X_b g_{RL}}{(M_{\tilde
t_1}^2-M_{\tilde t_2}^2)}\left(
\frac{f_6(R_{t_1},R_b)}{6 M_{\tilde t_1}^4} -
\frac{f_6(R_{t_2},R_b)}{6 M_{\tilde t_2}^4}\right)\nn \\ 
&-& \frac{m_t^2 X_t
g_{LR}}{M_{\tilde t_1}^2-M_{\tilde t_2}^2}\left(
\frac{f_{13}(R_{t_1},R_b)}{6 M_{\tilde t_1}^2} -
\frac{f_{10}(R_{t_1},R_b)}{3 M_{\tilde t_1}^2} -
\frac{f_{13}(R_{t_2},R_b)}{6 M_{\tilde t_2}^2}+
\frac{f_{10}(R_{t_2},R_b)}{3 M_{\tilde t_2}^2} \right)\nn \\
&-&\frac{m_b^2 X_b
g_{LR}}{12 M_{\tilde t_1}^4} f_{15}(R_{t_1},R_b)
-\frac{M_{\tilde g} g_{LR}}{24 M_{\tilde t_1}^4
}(m_t^2 f_2 + M_{H^+}^2 f_3 + m_b^2 f_4)(R_{t_1},R_b)\nn \\
&-&\left. \frac{M_{\tilde g} m_b^2 X_b^2 g_{LR}}{12 M_{\tilde
t_1}^6} f_8 (R_{t_1},R_b)- \frac{M_{\tilde g}(M_L^{'2}-M_R^{'2})g_{LR}}{12
M_{\tilde t_1}^4} f_6(R_{t_1},R_b)\right]\,,
\end{eqnarray}

\subsection*{{\bf{C.4}} Near-zero sbottom mixing and maximal stop mixing}

The results for $K_t$, $H_t$, $K_b$ and
$H_b$ up to order $M_{EW}^2/M_{SUSY}^2$ are:
\begin{eqnarray}
K_t &=& \frac{\alpha_s}{3 \pi}\,\left[ \frac{M_{\tilde g}X_t}
{M_S^2} f_1(R_t)
+\frac{1}{2}\, ln \frac{M_{\tilde b_1}^2}{M_S^2}-
g_1(R_{b_1}) + g_1(R_{t}) - g_2(R_{b_1}) 
+ g_2(R_{t}) \right. \nn \\
&-&\frac{m_b^2 X_b^2}{(M_{\tilde b_1}^2-M_{\tilde b_2}^2)^2}
\left(-\frac{1}{2}\, ln \frac{M_{\tilde b_2}^2}{M_{\tilde b_1}^2}+
g_1(R_{b_2}) -g_1(R_{b_1}) + g_2(R_{b_2}) - g_2(R_{b_1}) \right) \nn \\ 
&-&\frac{m_t^2}{12 M_S^2}(2f_2-f_3)(R_{t})
+\frac{m_t^2 X_t^2}{24 M_S^4}(4f_2 - 2f_3+f_4)(R_t)
-\frac{M_L^{2}-M_R^{2}}{12 M_S^2}(3f_1-f_2)(R_t) \nn \\
&+& \left. \frac{M_{\tilde g}m_b^2 X_b}{M_{\tilde b_1}^2-M_{\tilde b_2}^2}
\left(\frac{f_2(R_{b_1})}{3 M_{\tilde b_1}^2} - 
\frac{f_2(R_{b_2})}{3 M_{\tilde b_2}^2} \right) 
- \frac{m_b^2}{6 M_{\tilde b_1}^2}(2f_2-f_3)(R_{b_1})
- \frac{m_b^2}{12 M_{\tilde b_2}^2}(2f_2-f_3)(R_{b_2})\right]\,, \nn\\
H_t &=& - \frac{2\alpha_s}{3\pi} \frac{1}{m_t \cot\beta}\left[
-\frac{M_{\tilde g} g_{RL}}{2 M_S^2}f_1 (R_t,R_{b_1}) +\frac{m_t g_{LL}}{6
M_S^2} (2 f_{10} - f_{13})(R_t,R_{b_1})\right. \nn \\ 
&+&\frac{M_{\tilde g} g_{LL}}{6 M_S^4}m_t X_t f_5 (R_t,R_{b_1})
+\frac{m_b g_{RR}}{6 M_S^2} f_{13}(R_t,R_{b_2})\nn\\
&+& \frac{M_{\tilde g} m_b X_b}{M_{\tilde b_1}^2-M_{\tilde b_2}^2} 
\left(g_{RR}-\frac{m_b X_b}{M_{\tilde b_1}^2-M_{\tilde b_2}^2}g_{RL}\right)
\left(\frac{f_1 (R_{t},R_{b_2})}{2M_S^2}-
\frac{f_1(R_{t},R_{b_1})}{2M_S^2}\right) \nn \\
&-& \frac{m_t m_b X_b g_{LR}}{6 M_S^2
(M_{\tilde b_1}^2-M_{\tilde b_2}^2)} \left(f_{13}(R_t,R_{b_1}) -2 f_{10}(R_t,R_{b_1})-f_{13}(R_t,R_{b_2})+2f_{10}(R_t,R_{b_2})\right)\nn \\
&-&\frac{m_t m_b X_t g_{LR}}{24 M_S^4} f_{14}(R_t,R_{b_2})
+\frac{M_{\tilde g} m_t m_b X_t X_b g_{LR}}{6 M_S^4(M_{\tilde b_1}^2-M_{\tilde
    b_2}^2)}\left(f_5(R_t,R_{b_1})-f_5(R_t,R_{b_2})\right)\nn\\
&-& \frac{m_t^2 X_t g_{RL}}{24 M_S^4}
\left(3 f_{11}(R_t,R_{b_1})-f_{14}(R_t,R_{b_1})\right)\nn\\
&-& \frac{m_b^2 X_b
g_{RL}}{6 M_S^2(M_{\tilde b_1}^2-M_{\tilde b_2}^2)}
\left(f_{13}(R_t,R_{b_2})-f_{13}(R_t,R_{b_1})\right)\nn\\
&-&\frac{M_{\tilde g} g_{RL}}{24 M_S^4
}(m_t^2 f_2 + M_{H^+}^2 f_3 + m_b^2 f_4)(R_t,R_{b_1})-
\frac{M_{\tilde g} m_t^2 X_t^2 g_{RL}}{12 M_S^6}f_7 (R_t,R_{b_1})\nn \\
&-&\left. \frac{M_{\tilde g}(M_L^2-M_R^2)g_{RL}}{12
M_S^4} f_5 (R_t,R_{b_1})\right]\, \nn\\
K_b &=&  \frac{\alpha_s}{3 \pi}\,\left[ - \frac{2 M_{\tilde g}
X_b}{M_{\tilde b_1}^2-M_{\tilde b_2}^2} \left( ln\frac{M_{\tilde
b_2}^2}{M_{\tilde b_1}^2}+g_1 (R_{b_1}) - g_1 (R_{b_2}) \right) 
-\frac{1}{2}\, ln \frac{M_{\tilde b_1}^2}{M_S^2}+
g_1(R_{b_1}) \right. \nn \\
&-& g_1(R_{t}) + g_2(R_{b_1}) 
- g_2(R_{t}) +\frac{m_b^2 X_b^2}{(M_{\tilde b_1}^2-M_{\tilde b_2}^2)^2}
\left(-\frac{1}{2}\, ln \frac{M_{\tilde b_2}^2}{M_{\tilde b_1}^2}+
g_1(R_{b_2}) -g_1(R_{b_1})\right. \nn \\
 &+& \left.  g_2(R_{b_2}) 
- g_2(R_{b_1}) \right) 
-\frac{m_b^2}{12 M_{\tilde b_2}^2}(2f_2-f_3)(R_{b_2})\nn \\
&-&\frac{m_t^2}{4 M_S^2}(2f_2-f_3)(R_{t})
-\frac{m_t^2 X_t^2}{24 M_S^4}(4f_2 -2f_3+f_4))(R_t)
+\frac{M_L^{2}-M_R^{2}}{12 M_S^2}(3f_1-f_2)(R_t) \nn \\
&-& \left. \frac{M_{\tilde g}m_t^2 X_t}{6 M_S^4}
(2f_3-f_4)(R_t)\right]\,, \nn\\
H_b &=& - \frac{2\alpha_s}{3\pi} \frac{1}{m_b \tan\beta}\left[
-\frac{M_{\tilde g} g_{LR}}{2 M_S^2}f_1 (R_t,R_{b_2}) +\frac{m_t g_{RR}}{6
M_S^2} (2 f_{10} - f_{13})(R_t,R_{b_2})\right. \nn \\ 
&+&\frac{M_{\tilde g} g_{RR}}{6 M_S^4}m_t X_t f_5 (R_t,R_{b_2})+
\frac{m_b g_{LL}}{6 M_S^2} f_{13}(R_t,R_{b_1}) \nn\\ 
&+& \frac{M_{\tilde g} m_b X_b}{M_{\tilde b_1}^2-M_{\tilde b_2}^2} 
\left(g_{LL}+\frac{m_b X_b}{M_{\tilde b_1}^2-M_{\tilde b_2}^2}g_{LR}\right)
\left(\frac{f_1 (R_{t},R_{b_2})}{2M_S^2}-
\frac{f_1(R_{t},R_{b_1})}{2M_S^2}\right) \nn \\
&-& \frac{m_t m_b X_b g_{RL}}{6 M_S^2
(M_{\tilde b_1}^2-M_{\tilde b_2}^2)} \left(f_{13}(R_t,R_{b_1}) -2 f_{10}(R_t,R_{b_1})-f_{13}(R_t,R_{b_2})+2f_{10}(R_t,R_{b_2})\right)\nn \\
&-&\frac{m_t m_b X_t g_{RL}}{24 M_S^4} f_{14}(R_t,R_{b_1})
+\frac{M_{\tilde g} m_t m_b X_t X_b g_{RL}}{6 M_S^4 (M_{\tilde b_1}^2-M_{\tilde
    b_2}^2)}\left(f_5(R_t,R_{b_1})-f_5(R_t,R_{b_2})\right)\nn\\
&-& \frac{m_t^2 X_t
g_{LR}}{24 M_S^4}\left(3 f_{11}(R_t,R_{b_2})-f_{14}(R_t,R_{b_2})\right) \nn \\
&-&\left.\frac{m_b^2 X_b
g_{LR}}{6 M_S^2(M_{\tilde b_1}^2-M_{\tilde b_2}^2)}\left(f_{13}(R_t,R_{b_2})-f_{13}(R_t,R_{b_1})\right)\right. \nn \\
&-&\frac{M_{\tilde g} g_{LR}}{24 M_S^4
}(m_t^2 f_2 + M_{H^+}^2 f_3 + m_b^2 f_4)(R_t,R_{b_2})
-\frac{M_{\tilde g} m_t^2 X_t^2 g_{LR}}{12 M_S^6} f_7 (R_t,R_{b_2})\nn \\
&+&\left. \frac{M_{\tilde g}(M_L^2-M_R^2)g_{LR}}{12 M_S^4} f_5(R_t,R_{b_2})  
\right]\,. 
\end{eqnarray}

\vspace{0.4cm}

\begingroup\raggedright\endgroup

\end{document}